\documentclass[journal]{IEEEtran}
\usepackage{cite}
\usepackage{amsmath}
\usepackage{amsfonts}
\usepackage{url}
\usepackage{graphicx}
\usepackage{subfigure}
\usepackage{caption}
 \usepackage{epstopdf}
\usepackage{multirow}
\usepackage{color,xcolor}
\usepackage{cite}
\usepackage{CJK}
\usepackage{comment}
\usepackage{amsthm} % 如果要使用proof语句就必须要引入这个包

\usepackage{verbatim}
\usepackage{bbding}

\usepackage[pagebackref=true,breaklinks=true,colorlinks,bookmarks=false]{hyperref}
\newcommand{\R}{\ensuremath{\mathbb{R}}}
\newcommand{\N}{\ensuremath{\mathbb{N}}}

\begin{document}
\captionsetup{labelformat=default,labelsep=space}
\title{Once-Training-All-Fine: No-Reference Point Cloud Quality Assessment via Domain-relevance Degradation Description}
\author{Yipeng Liu, Qi Yang, Yujie Zhang, Yiling Xu, Le Yang, Xiaozhong Xu, Shan Liu\thanks{This paper is supported in part by National Natural Science Foundation of China (62371290), National Key R\&D Program of China (2024YFB2907204), the Fundamental Research Funds for the Central Universities of China, and STCSM under Grant (22DZ2229005). The corresponding author is Yiling Xu(e-mail: yl.xu@sjtu.edu.cn).}
\thanks{Y. Liu, Y. Zhang and Y. Xu are from Cooperative Medianet Innovation Center, Shanghai Jiaotong University, Shanghai, 200240, China, (e-mail: liuyipeng@sjtu.edu.cn, yujie19981026@sjtu.edu.cn, yl.xu@sjtu.edu.cn)}
\thanks{Q. Yang is from University of Missouri-Kansas City, Missouri, US, (e-mail: qiyang@umkc.edu)}
\thanks{L. Yang is from the Department of electrical and computer engineering, University of Canterbury, Christchurch 8041, New Zealand, (e-mail: le.yang@canterbury.ac.nz)}
\thanks{X. Xu, S. Liu are from Media Lab, Tencent, Shenzhen, China, (e-mail: xiaozhongxu@tencent.com, shanl@tencent.com)}
}
\IEEEtitleabstractindextext{
\begin{abstract}

The visual quality of point clouds plays a crucial role in the development and broadcasting of immersive media. Therefore, investigating point cloud quality assessment (PCQA) is instrumental in facilitating immersive media applications, including virtual reality and augmented reality applications. Considering reference point clouds are not available in many cases, no-reference (NR) metrics have become a research hotspot. Existing NR methods suffer from difficult training. To address this shortcoming, we propose a novel NR-PCQA method, Point Cloud Quality Assessment via Domain-relevance Degradation Description (D$^3$-PCQA). First, we demonstrate our model's interpretability by deriving the function of each module using a kernelized ridge regression model. Specifically, quality assessment can be characterized as a leap from the scattered perceptual domain (reflecting subjective perception) to the ordered quality domain (reflecting mean opinion score). Second, to reduce the significant domain discrepancy, we establish an intermediate domain, the description domain, based on insights from the human visual system (HVS), by considering the domain relevance among samples located in the perception domain and learning a structured latent space. The anchor features derived from the learned latent space are generated as cross-domain auxiliary information to promote domain transformation. Furthermore, the newly established description domain decomposes the NR-PCQA problem into two relevant stages. These stages include a classification stage that gives the degradation descriptions to point clouds and a regression stage to determine the confidence degrees of descriptions, providing a semantic explanation for the predicted quality scores. Experimental results demonstrate that D$^3$-PCQA exhibits robust performance and outstanding generalization on several publicly available datasets.

\end{abstract}
}
\maketitle
\IEEEdisplaynontitleabstractindextext
\IEEEpeerreviewmaketitle
\section{Introduction}\label{sec:intro}

\begin{figure}[t]
\centering
\includegraphics[width=1\linewidth]{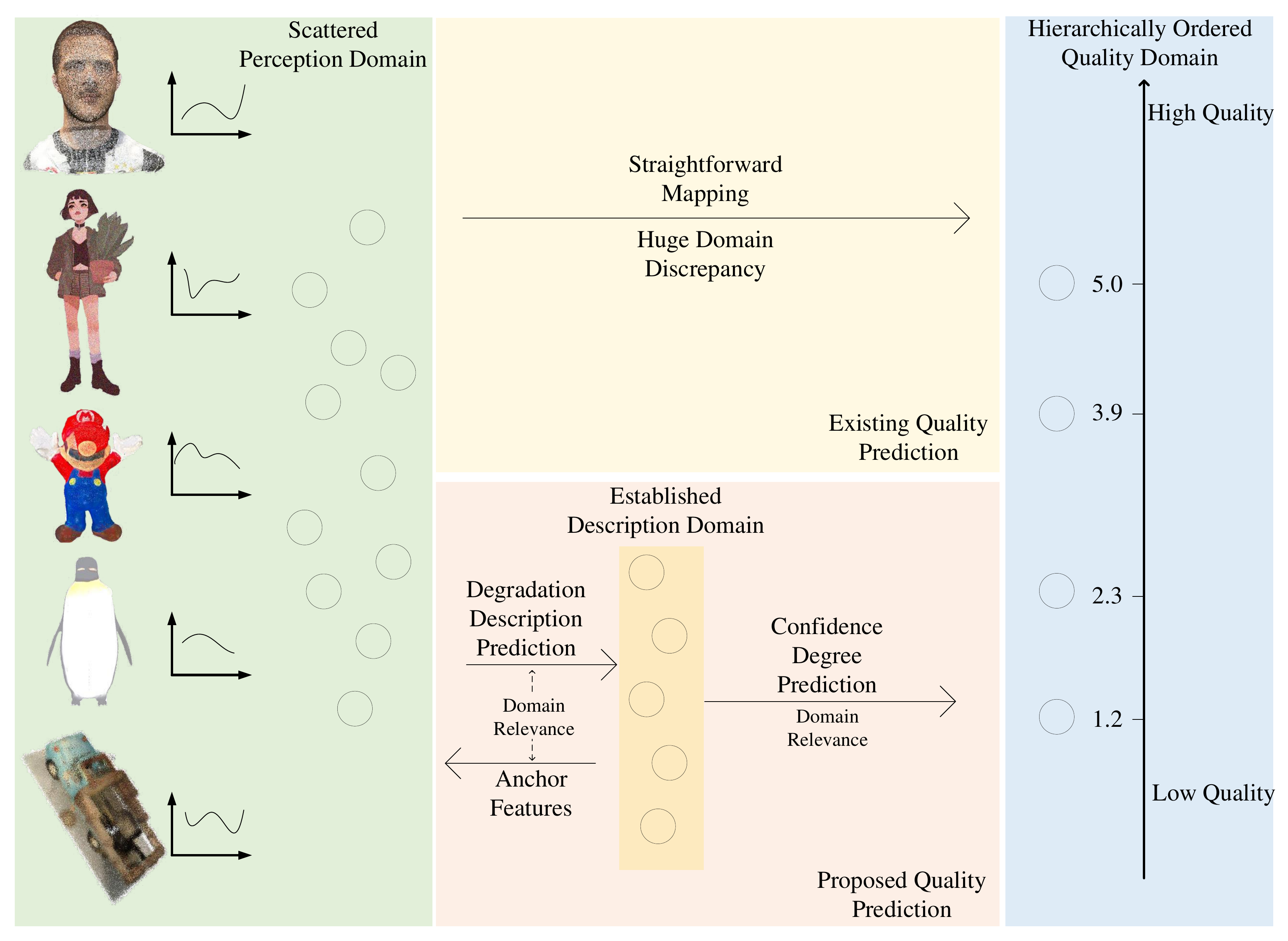}
\caption{Comparison between existing NR-PCQA methods and the proposed method. The huge discrepancy between the perception domain and quality domain poses a challenge in establishing a reliable mapping. To address this issue, the proposed method introduces a new description domain by considering the relevance among samples located in the perception domain, which helps to narrow the domain discrepancy. }
\label{fig:motivation}
\end{figure}

With the rapid advancement of the “meta-universe” and the increasing capabilities of 5G networks, the forms of 3D immersive media services have undergone continuous innovation~\cite{tbc5}. Among these emerging media forms, point clouds enable the representation and rendering of highly detailed 3D scenes, offering immersive and interactive experiences that are gaining significant attention in various broadcasting applications~\cite{Chen2021driving, Cheng2020medicalPC}. A point cloud is a collection of non-uniformly scattered 3D points that may suffer from impairments in both geometry and attributes (\textit{e.g.,} color) during processing, resulting in perceptual degradation. To facilitate immersive visual tasks (\textit{e.g.,} virtual reality~\cite{Alexiou2020PointXR} and augmented reality~\cite{AR1}), point cloud quality assessment (PCQA) has gained significant attention among researchers. PCQA can be achieved through subjective experiments or objective metrics. However, although subjective experiments can provide the ultimate prediction, they can be expensive in terms of time, cost and testing conditions~\cite{Mantiuk2012QAsurvey}. Therefore, designing effective objective metrics has become a hotspot in recent research. Objective metrics can be categorized as full-reference (FR), reduced-reference (RR) and no-reference (NR) methods. FR and RR metrics require the entire original samples or partial features as a reference, which may not be readily available in most scenarios. Thus, we focus on NR metrics as they are designed for scenarios where the high-quality original point cloud is not available.

\subsection{Motivation}

Current NR metrics are mostly based on deep learning. The common strategy is to use well-designed deep neural networks to map the input point clouds into the feature space, and then regress the final scores using the obtained latent features~\cite{Yang2019DNNIQA}, which can be formulated as $q = f(\phi(x))$, where $x$ and $q$ represent the input point cloud and final objective score, and $\phi(\cdot)$ and $f(\cdot)$ represent the feature extraction and quality mapping operation. However, the performance of this architecture suffers from difficult training and is far from satisfactory. The main reason is that this paradigm ignores or weakens some important intermediate developments of subjective evaluation.

The process of quality assessment entails the transformation between different domains. Based on the study of the human visual system (HVS), we know that vision begins with the cone cells in the retina. The layer of nerve cells transmits visual signals to the brain, culminating in the generation of the quality perception~\cite{Simoncelli2001neur, Felsen2005neur}.  We define the distribution of immediately perceived visual stimuli as the \textbf{perception domain}, and the distribution of final subjective quality scores as the \textbf{quality domain}.
The perception domain scatters high-dimensionally due to the presence of massive visual information, while the quality domain is a hierarchically ordered space with a limited range depending on the settings of the subjective experiment~\cite{Liu2022LSPCQA, Yang2020TMM3DTO2D}. Objective methods align with subjective perception in their shared objective of achieving domain transformation. However, NR-PCQA methods face a first problem stemming from the scattered distribution of training data in the perception domain, which hinders model fitting. Point clouds exhibit greater complexity in terms of geometry and attributes compared with images. Nevertheless, existing PCQA datasets, such as PointXR~\cite{Alexiou2020PointXR}, IRPC~\cite{Javaheri2019IRPC}, ICIP2020~\cite{Perry2020ICIP20}, M-PCCD~\cite{Alexiou2019MPCCD}, SJTU-PCQA~\cite{Yang2020TMM3DTO2D}, and WPC~\cite{Su2019WPC,Liu2022WPC} typically consist of merely a few hundred distorted point clouds, resulting in only a few dozen samples for each distortion type. Considering the significant variability in content and distortion types, the available distorted point clouds are usually dispersedly scattered in the high-dimensional perception domain, resulting in a huge domain discrepancy with the ordered quality domain. The huge domain discrepancy poses challenges for domain transformation.

The second problem arises from the disparity between current NR-PCQA frameworks and HVS mechanism. The existing NR-PCQA methods attempt to establish a direct mapping between the scattered perception domain and ordered quality domain, which is contradictory to subjective observation \cite{liu2022Transfer}.  In fact, HVS requires a training session that signifies the intrinsic relevance within the perception domain~\cite{Yang2023TMD} as prior knowledge to facilitate domain transformation.  Furthermore, the discrete human learning and categorization model makes the HVS first acquire discrete degradation descriptions~\cite{Kulkarni2013SemanDesp, Aitkin2009Discretization} by retrieving the prior knowledge \cite{Buckner1996template}, and then use quantitative value to represent the perceiving judgment regarding the degradation \cite{BT500}. Overall, during the subjective evaluation process, HVS naturally establishes an intermediate domain related to the discrete degradation descriptions. However, most current objective methods neglect the presence of this intermediate domain.

To address the above two problems, we intend to achieve a progressive quality prediction. Considering the robustness of subjective assessment, the involvement of the intermediate domain can effectively reduce the difficulty associated with domain transformation. Therefore, we define a \textbf{description domain} located between the perception domain and quality domain (seeing Fig. \ref{fig:motivation}), which pertains to the feature distribution for discrete degradation descriptions. The distillation of perception domain features, which corresponds to the formation of prior knowledge of HVS, in fact reinforces the commonality of samples with the same degradation description and gives rise to the generation of the description domain. The elements in the quality domain serve as the fine-grained expression of the description domain features. Merging the description domain into the implementation of neural networks, we can bridge the perceptual domain and quality domain in a progressive manner, therefore achieving a smoother domain transformation.

% The third problem is that most NR-PCQA methods utilize a projection-based backbone which converts point clouds to images with predefined resolutions. This choice is made due to its lower memory usage and faster inference speed when compared with consuming 3D schemes~\cite{Liu2022LSPCQA}. However, these projection images exhibit different perceptions compared with original point clouds, primarily due to information loss and masked distortions, limiting the performance potential of projection-based methods.  

\subsection{Our Approach}
\label{ourapproach}

We propose a novel NR-PCQA method called Point Cloud Quality Assessment via Domain-relevance Degradation Description (D$^3$-PCQA) regarding the above problems. The core idea of D$^3$-PCQA is to establish the ignored description domain. Given the widespread recognition of the 5-level quality description \cite{Li2022REQA,Ou2023ICCV}, we employ BT.500~\cite{BT500} for modeling as illustrated in Table \ref{Score}. In this case, the simulated description domain decomposes the predicted quality into an integer fundamental quality level referring to five discrete degradation descriptions, and a decimal confidence degree derived from preference confidences. The generation of quality is designated as Stage-1, and the calculation of confidence degree is identified as Stage-2.

First, to implement the proposed method, the training set is divided into a support set and a query set. The support set which is divided into five quality levels according to Table \ref{Score} is used to establish the description domain, and the query set is used to train the quality regression network which will be introduced in the following part.

In Stage-1, we predict the quality level by learning a structured latent space, which reproduces the function of the description domain. We employ ResNet50 as a basic backbone to generate the perception domain features with a HVS-based projection method which represents point clouds by emulating the effect of observation distance on HVS. To establish the description domain, the perception domain features are first disentangled into the domain-invariant features by an attention transformer network. Then the required structured latent space is learned by promoting the regular distribution of the disentangled features using the support set with a series of constraints. This process exploits the intrinsic relevance among different samples of the same quality level (namely intra-domain relevance). The clustering centers corresponding to each quality level, which contain the representative feature for the local description domain, can be used as the \textbf{anchor features} to promote domain transformation, simulating the learning of prior knowledge. Finally, inspired by the retrieval of prior knowledge in HVS, we measure the relevance between the query set samples and the five anchor features (namely inter-domain relevance) and map it into the quality level by a proposed classification network.

In Stage-2, we assess the confidence degree of the assigned quality level by measuring the feature relevance within the local description domain corresponding to a specific quality level. The measurement and mapping of feature relevance between the query set sample and the support set samples of determined quality level (which is yet another instance of intra-domain relevance) are approached as a regression problem, leading to the determination of the confidence degree. Finally, we obtain the final quality score by combining the quality level generated in Stage-1 with the confidence degree derived from Stage-2.

To elucidate the roles and functionalities of individual modules within the artificial neural network, we derive the pipeline of each module from a kernelized ridge regression model.

% Table generated by Excel2LaTeX from sheet 'Sheet3'
\begin{table}[t]
  \centering
  \caption{Five-grade degradation description defined by BT.500~\cite{BT500}.}
  \begin{footnotesize}
    \begin{tabular}{c|l}
    \hline
    MOS    & Degradation Description \\
    \hline
    5     & The distortion is almost imperceptible \\
    \hline
    4     & The distortion is perceptible but not annoying \\
    \hline
    3     & The distortion is slightly annoying \\
    \hline
    2     & The distortion is annoying \\
    \hline
    1     & The distortion is seriously annoying \\
    \hline
    \end{tabular}%
    \end{footnotesize}
  \label{Score}%
\end{table}%

\subsection{Contributions}

The contributions of this paper are summarized as follows:

\begin{itemize}
\item We propose a novel NR-PCQA method called D$^3$-PCQA to emulate the discrete learning and categorization mechanisms in HVS.

\item We greatly reduce training difficulty for quality prediction by exploiting the intra-domain relevance in the perception domain to establish the description domain.

\item  The proposed D$^3$-PCQA shows reliable performance and outstanding generalization ability. 

\end{itemize}

The rest of this paper is organized as follows. The related work is surveyed in Section \ref{sec:relatedwork}. Section \ref{sec:Reformulation} derives the functionality of each module from the solutions to a kernelized ridge regression problem. Section \ref{sec:network} presents the network implementation of the proposed D$^3$-PCQA, with its performance evaluation given in Section \ref{sec:experiments}. Finally, the conclusion is drawn in Section \ref{sec:conclusion}.

\section{Related Work}
\label{sec:relatedwork}

This section reviews existing PCQA metrics.

For FR-PCQA, Moving Picture Experts Group (MPEG) has applied point-to-point (p2point)~\cite{cignoni1998metro}, point-to-plane (p2plane)~\cite{Tian2017Evaluation} and PSNRyuv~\cite{MPEGSoft} in point cloud compression (PCC) standardization \cite{tbc12}. Other point-wise metrics, such as those proposed in ~\cite{alexiou2018pointt} and~\cite{javaheri2020haus}, have also been made available. Considering the geometry and color attributes simultaneously, Meynet \emph{et al.}~\cite{meynet2019pcmsdm} proposed a metric that pools curvature statistics and color lightness together via optimally-weighted linear combination~\cite{meynet2020pcmd}. Viola \emph{et al.}~\cite{viol2020acolor} suggested quantifying point cloud quality using color histograms.  Alexiou \emph{et al.}~\cite{alexiou2020TowardsStructural} incorporated four types of point cloud attributes into the form of SSIM~\cite{wang2004image}. Yang \emph{et al.}~\cite{yang2020graphsim} utilized color gradient to estimate point cloud quality based on graph signal processing. Zhang \emph{et al.}~\cite{Zhang2021MSGraphSIM} further improved ~\cite{yang2020graphsim} by combining the multi-scale representation of MS-SSIM~\cite{Wang2003MSSSIM}. Javaheri \emph{et al.}~\cite{javaheri2021PTD} developed a point-to-distribution metric to measure point cloud quality. Another approach for FR-PCQA is to project the 3D point cloud onto a number of 2D planes and then represent point cloud quality using the weighted indices of these image planes. Torlig\emph{et al.}~\cite{torlig2018novel} proposed real-time voxelization and projection techniques to present point clouds and evaluated IQA metrics for PCQA. Alexiou \emph{et al.}~\cite{alexiou2018pointt} measured the distortion using the angles between tangent planes perpendicular to point normals. Yang \emph{et al.}~\cite{Yang2020TMM3DTO2D} combined global and local features of projection planes to estimate point cloud quality. Xu \emph{et al.} \cite{tbc3} and Yang \emph{et al.} \cite{yang2021mped} introduced the point potential energy into the point cloud quality modeling. Javaheri \emph{et al.}~\cite{javaheri2021JPC} proposed a joint geometry and color projection and applied 2D quality metrics to reflect point cloud quality. These metrics further improve the performance of quality assessment by considering HVS-related features.

For RR-PCQA, Viola \emph{et al.}~\cite{Irene2020RR2} inferred point cloud quality using statistical information of geometry, color and normal vector. Q. Liu \emph{et al.}~\cite{Liu2021RR, Liu2022add2}, Su \emph{et al.} \cite{Su2023Bitstream} and Y. Liu \emph{et al.}~\cite{liu2022VCIP} estimated quality using compression parameters to guide PCC strategy with certain rate constraints \cite{tbc11}. Su \emph{et al.}~\cite{Su2024SVM} introduced the least absolute shrinkage and selection operator to leverage compression, geometry, normal, curvature, and luminance features for quality prediction. These methods still rely on reference features. However, acquiring the reference version can be challenging in practical scenarios, prompting the development of NR metrics.

Like in FR-PCQA, NR metrics can be performed either over the 2D projection of point clouds or directly on the raw data. For methods conducted over point cloud projection, Tao \emph{et al.}~\cite{Tao2021PMBVQA} employed multi-scale feature fusion to predict the quality of point clouds. Liu \emph{et al.}~\cite{Liu2021PQANet} proposed to leverage distortion classification information as an auxiliary feature to assist in the training of the network. Yang \emph{et al.}~\cite{Yang2021ITPCQA} bridged conventional images and point cloud projection via domain adaptation to expand the scale of trainable point cloud data. Fan \emph{et al.}~\cite{Fan2022Video} and Zhang \emph{et al.}~\cite{Zhang2022Video} integrated the point cloud projection into a video, followed by the utilization of video quality assessment methods for the purpose of evaluating the quality of point clouds. For methods over raw 3D data, Liu \emph{et al.}~\cite{Liu2022LSPCQA} adopted an end-to-end sparse convolution network to learn the quality representation of point clouds. Shan \emph{et al.}~\cite{shan2022GPANet} extracted anti-perturbation features for point clouds using a graph neural network. Wu \emph{et al.}~\cite{tbc4} extracted global and local quality features based on point cloud graph. Liang \emph{et al.}~\cite{tbc2} fused hierarchical features and hand-crafted features as the point cloud quality representation. In addition, other algorithms have been developed that leverage both point cloud projection and raw 3D data to extract integrated features, as exemplified by the work of Zhang et al. \cite{Zhang2022MMPCQA}. These methods adopt a uniform architecture of feature extraction using a range of techniques, followed by regression into score values that fail to accurately emulate the intricate human visual mechanism.  In this work, we aim to introduce a fresh approach to the NR-PCQA problem by mapping extracted features to align with human perception. 

\section{Problem Formulation }
\label{sec:Reformulation}

In this section, we illustrate each module of D$^3$-PCQA from the perspective of a kernelized ridge regression to better understand the architecture of the proposed method.

Given the training samples $\{ {x_i},{y_i}\} _{i = 1}^N$ where $x_i$ is the $i$-th distorted sample and $y_i$ is the \textit{continuous }quality score, the existing NR-PCQA can be formulated as a generalized linear ridge regression problem:
\begin{align}
\label{formu1}
w = \arg \mathop {\min }\limits_w \sum\limits_{i = 1}^N {{{({y_i} - w^T\varphi ({x_i}))}^2} + \frac{\lambda}{2} {{\left\| w \right\|}^2}},
\end{align}
where $\varphi(\cdot)$ represents the nonlinear feature mapping, $w$ is the regression vector, $N$ is the total number of training samples, and $\lambda$ signifies the trade-off parameter for the regularization term. The regularization term indicates the lightweight tendency of existing NR-PCQA methods. 

Assume that participants’ preferences for degradation descriptions of the same sample are distributed within the same quality level, in order to establish a new intermediate domain between the perception domain and quality domain, we may express the label $y$ and response $r=w_{}^T\varphi ({x})$ in (\ref{formu1}) as the summation of an integer quality level and decimal degree of confidence. As a result, we can decompose the original NR-PCQA task aimed at producing a continuous objective score into the combination of a multi-class classification problem that finds the integer quality level corresponding to the quantification of degradation description and a regression problem that generates the degree of preference confidence. Mathematically, we have
\begin{align}
{y} & = {y_{L}} + {y_{R}}, \\
{r} & = w_{}^T\varphi ({x}) = {r_{L}} + {r_{R}},
\end{align}
where $y_{L}$ and $r_{L}$ indicate the \textit{integer} quality level of $y$ and $r$, and $y_{R}$ and $r_{R}$ represent the \textit{decimal} degree of confidence of $y$ and $r$. 

\subsection{Stage-1: Degradation Description Prediction}
The discrete degradation description defined by BT.500~\cite{BT500} (Table \ref{Score}) of a testing sample $x_i$ can be determined by classifying $y_{i,L}\in \{1,2,3,4,5\}$. In order to establish the description domain corresponding to $y_{i,L}$, the training samples are grouped according to the distribution of their normalized quality scores. Correspondingly, the summands in \eqref{formu1} can be organized in groups and the quality level can then be determined by concatenated ridge regression models:
\begin{align}
\label{formu2}
w_L = \arg \mathop {\min }\limits_{w_L} \sum\limits_{j{ = }1}^{N/K} {[ {\sum\limits_{i{ = }1}^K {{{({y_{i,j,L}} - w_{j,L}^T\varphi ({x_{i,j}}))}^2}}  + \frac{\lambda}{2}{{\left\| {{w_{j,L}}} \right\|}^2}} ]},
\end{align}
where $x_{i,j}$ signifies the $i$-th samples in the $j$-th quality level group, and $K$ is the number of training samples for each quality level. $w_L$ is defined as $w_L = \{w_{j,L}\}_{j=1}^{N/K}$. Note that in (\ref{formu2}), the regressand has been changed to $y_{i,j,L}$, the class label corresponding to the degradation description of $y_{i,j}$. In other words, it is now a multi-class classification problem and we aim at identifying the sample's quality level (in terms of $y_{i,j,L}\in \{1,2,3,4,5\}$). The regression vector for the $j$-th quality level is from the representer theorem \cite{Scholkopf2001representer},  
\begin{align}
\label{formu3}
{w_{j,L}} = \sum\limits_{i = 1}^K {\alpha _{i,j,L}^{}\varphi ({x_{i,j}})},
\end{align}
where $x_{i,j}$ is the $i$-th distorted sample with its quality level equal to $j$. Using \eqref{formu3} transforms \eqref{formu2} into
\begin{align}
\label{formu5}
\begin{split}
\alpha_L  = \arg \mathop {\min }\limits_{\alpha_L}  \sum\limits_{j = 1}^{N/K} \bigg[\sum\limits_{i = 1}^K {{{({y_{i,j,L}} - \sum\limits_{k = 1}^K {{\alpha _{k,j,L}}{\varphi^T {({x_{k,j}})}\varphi ({x_{i,j}})}} )^2}}} \\
 + \frac{\lambda }{2}\sum\limits_{i = 1,k = 1}^K {{\alpha _{i,j,L}}{\alpha _{k,j,L}}{\varphi^T {({x_{k,j}})}\varphi ({x_{i,j}})}} \bigg],
\end{split}
\end{align}
where $w_{j,L} = \sum\limits_{k = 1}^K {{\alpha _{k,j,L}}\varphi ({x_{k,j}})}$ can be considered as the common feature extracted from the samples with the $j$-th quality level, which can characterize a representative feature for the local description domain corresponding to a specific degradation description. Let 
\begin{align}
\label{formu8}
\varphi_j ({x_{m}}) = w_{j,L} = \sum\limits_{k = 1}^K {{\alpha _{k,j,L}}\varphi ({x_{k,j}})},
\end{align}
which is called the \textbf{anchor feature} deriving from the intra-domain relevance of a specific quality level. 

Let ${k(x_{k,j},x_{i,j})} = \varphi^T {({x_{k,j}})}\varphi (x_{i,j})$ be the positive definite kernel function. The response can be rewritten as
\begin{align}
\label{formu2-a}
{r_{i,j,L}} = \varphi^T_j {({x_{m}})}\varphi ({x_{i,j}}) = k_j({x_{m}},{x_{i}}),
\end{align}
which now measures the inter-domain relevance between an input distorted sample $x_i$ and the anchor feature $\varphi_j (x_m)$ for $j$-th quality level. Therefore, we indeed utilize the relevance measurement to classify the degradation description for quality assessment.

% the anchor feature $\phi (x_m)$ and the feature of 

The kernel function $k_j({x_m},{x_i})$ in (\ref{formu2-a}) may be evaluated using the following quadratic form to further take into account the interaction between elements in the feature vectors $\varphi_j{(x_m^{})}$ and $\varphi (x_{i,j}^{})$:
\begin{align}
\label{formu2-5}
k'_{m,i,j} = \varphi^T_j {(x_m^{})}\beta_{j,L} \varphi (x_{i,j}^{}),
\end{align}
as long as the trainable matrix $\beta_{j,L}$ is at least positive semidefinite. In the alternating iteration method, the method in ~\cite{Higham1988semidefinite} can be invoked to approximate the estimated $\beta_{j,L}$ using the nearest (in terms of Frobenius norm $||\cdot||_F$) positive semidefinite matrix.

Substituting \eqref{formu2-5} into \eqref{formu5}, we obtain the following optimization problem
\begin{align}
\small
\label{formu6}
\begin{split}
\alpha_L ,\beta_L  = \arg \mathop {\min }\limits_{\alpha_L ,\beta_L}  \sum\limits_{j=1}^{N/K} \bigg[\sum\limits_{i=1}^K {{{({y_{i,j,L}} - \varphi^T_j {(x_m^{})}\beta_{j,L} \varphi (x_{i,j}^{}) )^2}}} \\
 + \frac{{{\lambda _1}}}{2}\sum\limits_{i = 1,k = 1}^K {{\alpha _{i,j,L}}{\alpha _{k,j,L}}{k'_{i,k,j}}}  + \frac{{{\lambda _2}}}{2}||\beta_{j,L}||^2_F\bigg ].
\end{split}
\end{align}

The solution to (\ref{formu6}) can be obtained by leveraging the anchor feature through evaluating $\varphi_j ({x_m}) = \sum\limits_{k = 1}^K {{\alpha _{k,j,L}}\varphi ({x_{k,j}})}$, $j=1,2,...,5$. For an unseen distorted sample $x_i$, its degradation description can then be determined using 
${r_{i,j,L}} = \varphi^T_j {(x_m^{})}\beta_{j,L} \varphi (x_i^{})$. The value of $j$ corresponding to the largest $r_{i,j,L}$ would be output as the quality level of $x_i$.

\subsection{Stage-2: Confidence Degree Prediction}
We can adopt the same nonlinear ridge regression framework in (\ref{formu6}) to map the coarse-grained quality level obtained in Stage-1 into the accurate quality score in the quality domain. Specifically, with slight abuse of notations, we aim at solving
\begin{align}
\scriptsize
\label{formu2-b}
\begin{split}
\alpha_R ,\beta_R  = \arg \mathop {\min }\limits_{\alpha_R ,\beta_R}  \sum\limits_{j = 1}^{N/K} \bigg [\sum\limits_{i = 1}^K {{{({y_{i,j,R}} - 
\sum\limits_{k = 1}^K {{\alpha _{k,j,R}}\varphi^T {(x_{k,j}^{})}\beta_{j,R} \varphi (x_{i,j}^{}) } )^2}}} \\
 + \frac{{{\lambda _1}}}{2}\sum\limits_{i = 1,k = 1}^K {{\alpha _{i,j,R}}{\alpha _{k,j,R}}{k'_{i,k,j}}}  + \frac{{{\lambda _2}}}{2}||\beta_{j,R}||^2_F \bigg ].
\end{split}
\end{align}
The regressand $y_{i,j,L}$, which is the class label for quality level prediction, is replaced with $y_{i,j,R}$, which is the decimal degree of confidence. Note that the main difference is that the anchor feature $\varphi_j ({x_m})$ for $j$-th quality level is not calculated first. Instead, the entire formula is employed to measure the intra-domain relevance between $\varphi (x_{i,j})$ and sample features of a specific quality level. This is because measuring feature relevance is critical for fine-tuning.

Since here both $\alpha_R$ and $\beta_{j,R}$ contribute to mapping the relevance, $\beta_{j,R}$ in (\ref{formu2-b}) is fixed to be identity matrix of an appropriate size to reduce the computational burden. We thus only need to find $\alpha_R$ such that the estimated decimal degree of confidence for an unseen sample $x_i$, whose quality level was found to be equal to $j$,  can be calculated using 
$\sum\limits_{k = 1}^K {{\alpha _{k,j,R}}\varphi^T {(x_{k,j}^{})}\varphi (x_{i}^{}) }$. Combining this result with the obtained quality level yields the final estimate of the continuous quality score for a testing sample.

\subsection{Functions of Derived Modules}
\begin{figure}[t]
\centering
\includegraphics[width=1\linewidth]{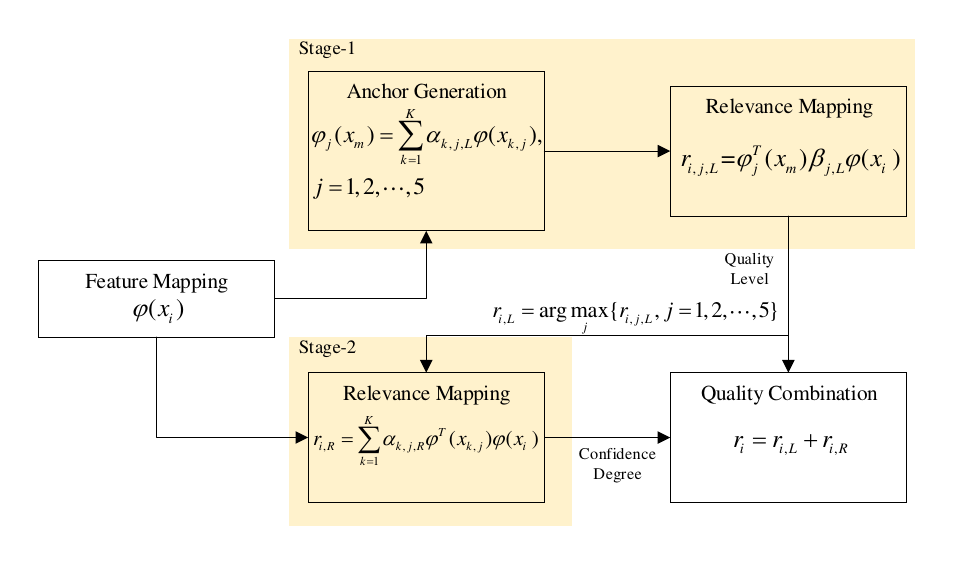}
\caption{Functions of derived modules. The derivation decomposes the conventional regression-based architecture into 2 relevant stages by establishing a new description domain, with the domain relevance being utilized for both domain establishment and domain transformation.}
\label{fig:coarse}
\end{figure}

Based on the above derivation, several modules can be identified from the response terms of (\ref{formu6}) and (\ref{formu2-b}), as illustrated in Fig. \ref{fig:coarse}. The proposed method includes the following modules:

\begin{itemize}
\item \textbf{Feature Extraction:} The original feature of sample $x_i$ is extracted from a feature extraction backbone, denoted as $\varphi(x_i)$.

\item \textbf{Anchor Feature Generation:} The description domain is established, where the \textit{anchor feature} for $j$-th quality level is generated by measuring the \textit{intra-domain relevance} among sample features in the perception domain, which is 
\begin{align}
\label{formu2-13}
\varphi_j ({x_m}) = \sum\limits_{k = 1}^K {{\alpha _{k,j,L}}\varphi ({x_{k,j}})}, j=1,2,...,5.
\end{align}

\item \textbf{Relevance Mapping:} In Stage-1, the quality level is determined by measuring the \textit{inter-domain relevance} between the sample feature and five anchor features, \textit{i.e.,} 
\begin{align}
\label{formu2-14}
{r_{i,L}}= \mathop {\arg\max \limits_j} \{\varphi^T_j {(x_m^{})}\beta_{j,L} \varphi (x_i^{}), j=1,2,...,5\}. 
\end{align}
In Stage-2, the confidence degree is determined by measuring the \textit{intra-domain relevance} among samples of the determined quality level, which is 
\begin{align}
\label{formu2-15}
{r_{i,R}} = \sum\limits_{k = 1}^K {{\alpha _{k,j,R}}\varphi^T {(x_{k,j}^{})}\varphi (x_i^{}) }.
\end{align}

\item \textbf{Quality Combination:} The quality level and the confidence degree are combined to generate the continuous quality score, given as ${r_i} = {r_{i,L}} + {r_{i,R}}$.

\end{itemize}

\begin{figure*}[ht]
\centering
\includegraphics[width=1\linewidth]{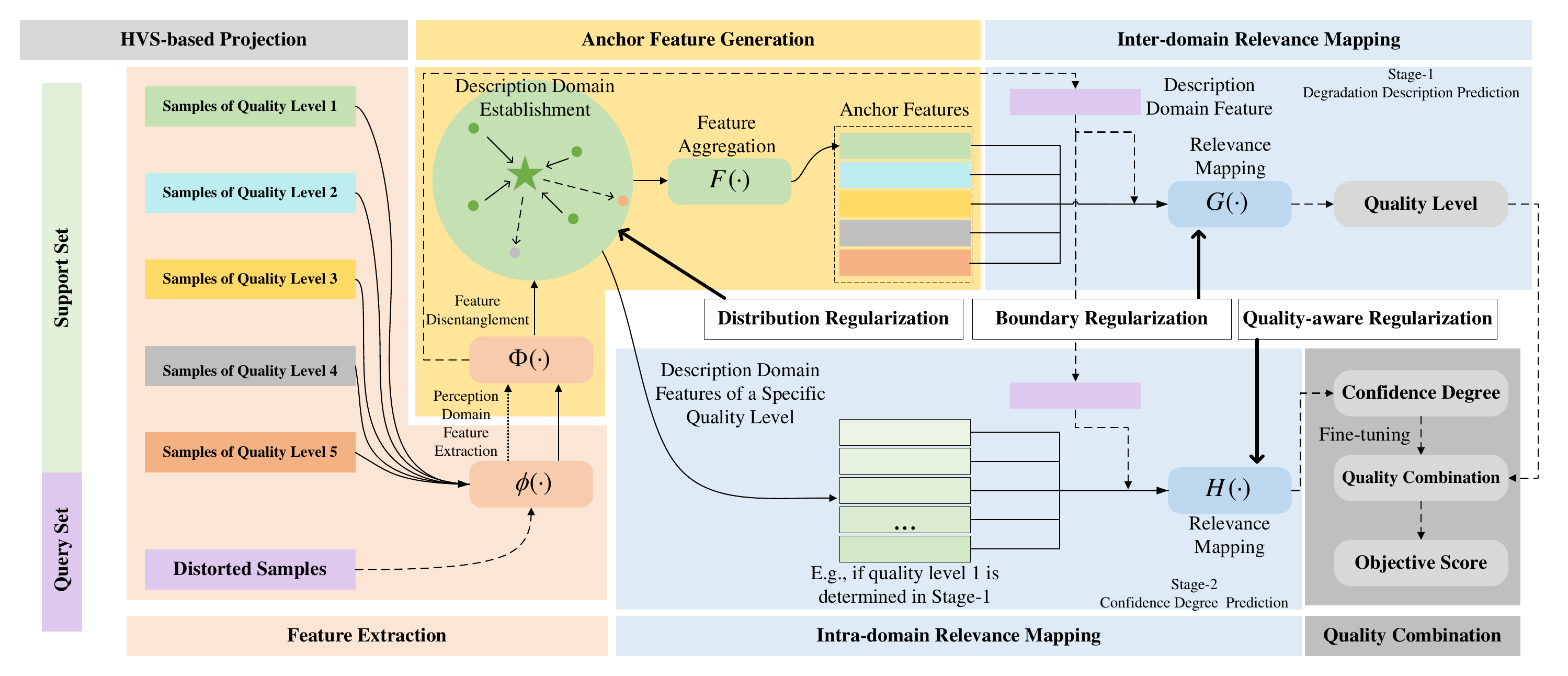}
\caption{Overall architecture of D$^3$-PCQA. D$^3$-PCQA is divided into 2 stages: degradation description prediction and confidence degree prediction. To implement D$^3$-PCQA, the training samples are partitioned into the support set and the query set, and the support set is further partitioned into five-grade impairment scales based on BT.500~\cite{BT500}. The realized lines indicate the feature flow of support set samples, while the dotted lines indicate the query set samples'. In Stage-1, the feature extraction module $\phi(\cdot)$ extracts the perception domain features, using the HVS-based projection, which are disentangled using the feature disentanglement module $\Phi ( \cdot )$. The description domain is established with a series of constraints based on the disentangled features in the support set, and then the anchor feature for each quality level is generated using the feature aggregation module $F( \cdot )$. The relevance mapping module $G(\cdot)$ solves the degradation description prediction as a classification problem by measuring the inter-domain relevance between the query set samples and five anchor features. In Stage-2, the relevance mapping module $H(\cdot)$ regresses the confidence degree based on the intra-domain relevance within the local description domain corresponding to the determined quality level. The quality combination module finally combines the quality level and the confidence degree to obtain the continuous quality score.}
\label{fig:framework}
\end{figure*}

\section{Network Implementation}
\label{sec:network}

We implement equivalent functionalities for \eqref{formu2-13}, \eqref{formu2-14} and \eqref{formu2-15} using neural networks, with slight deviations from Section \ref{sec:Reformulation}. These deviations are due to differences between the alternating iteration method and neural network implementations, as well as the realization of learning-based operations to contain non-linearities. The proposed D$^3$-PCQA framework, shown in Fig. \ref{fig:framework}, aims to capture previously ignored description domain information for performance improvement. It includes the feature extraction module ($\phi(\cdot)$), anchor feature generation module, relevance mapping module ($G( \cdot )$ and $H( \cdot )$), and quality combination module. The anchor feature generation module comprises the feature disentanglement ($\Phi ( \cdot )$), description domain establishment, and feature aggregation ($F( \cdot )$). Each module is discussed in detail in subsequent sections.

\begin{figure*}[t]
\centering
\includegraphics[width=1\linewidth]{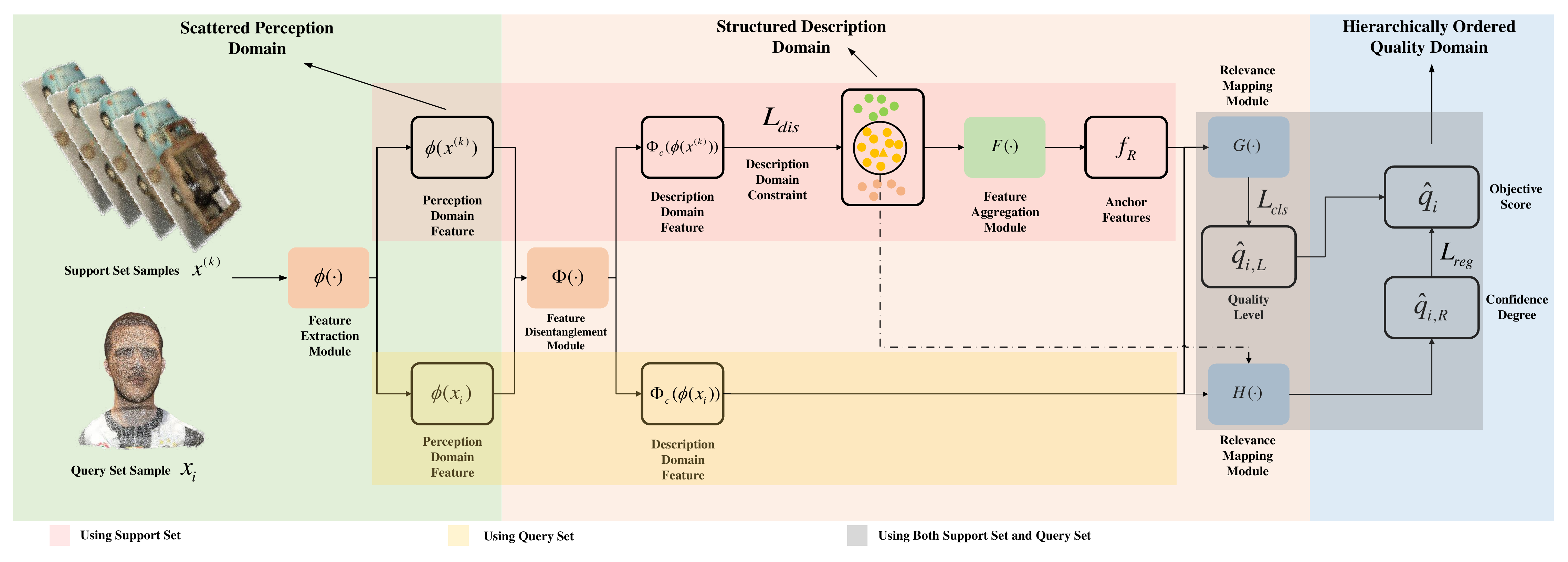}
\caption{Transformation between different domains of the output of each module. The module symbols and parameter variables are consistent with those in Fig. \ref{fig:framework} and in Section \ref{sec:network}. $\phi(\cdot)$ extracts the scattered perception domain features. $\Phi ( \cdot )$ isolates components related to the description domain from the perception domain features, and then the anchor feature generation by $F( \cdot )$ and the quality level prediction through $G( \cdot )$ can be performed in the description domain which is established with a series of constraints. Finally,  the confidence degree is predicted using $H( \cdot )$ to facilitate the transformation from the description domain to the quality domain. Besides, this schematic also shows the use of the support set and query set by different modules.}
\label{fig:domain}
\end{figure*}

Inspired by \cite{Hospedales2022meta,fewshot2018Sung}, we train the network by partitioning samples into support and query sets. The use of the support and query sets by different modules is shown in Fig. \ref{fig:domain}. Briefly, the support set establishes the description domain and generates the anchor features, while the query set trains the quality regression network.

Fig. \ref{fig:domain} illustrates the transformation between domains. The feature extraction module $\phi(\cdot)$ extracts perception domain features reflecting visual stimuli. The feature disentanglement $\Phi ( \cdot )$ isolates description-related components, enabling anchor feature generation ($F( \cdot )$) and quality level prediction ($G( \cdot )$) within the constrained description domain. Finally, $H( \cdot )$ predicts the confidence degree, bridging the description and quality domains.

To provide better illustration, we use the established description domain to decompose our proposed model into two stages, \textit{i.e.,} Stage-1 to predict the quality level and Stage-2 to generate the confidence degree. 

\subsection{Feature Extraction Module}
\label{sec:featureextraction}

In this work, we focus on the projection-based backbone to extract the perception domain features, which consumes less memory and has a faster inference speed compared with the 3D-based schemes. To enhance the 2D backbone, we incorporate some important visual characteristics that align projection images with human perception.

\subsubsection{Simulation of HVS Observation Mechanism}\label{sec: hvs_projection}

Information loss during immediate point cloud projection often stems from mismatches between point cloud and projection plane sizes. While existing methods like cropping \cite{Tao2021PMBVQA} and folding \cite{Quach2020folding} normalize input sizes, they distort whole perception and introduce perceptual distortions. To address this, we apply an HVS-inspired multi-scale point cloud representation. Specifically, projecting point clouds onto a small viewing window can be emulated as human perception of distant objects, where visual phenomena including scale reduction, detail loss, and blurring occur \cite{Wang2003MSSSIM,Zhang2021MSGraphSIM}. According to \cite{Zhang2021MSGraphSIM}, smaller-scale point clouds, resulting from long observation distances, still correlate with subjective scores. Thus, downscaling point clouds based on the HVS mechanism facilitates the generation of perceptually consistent projection images.

Our 2D representation approach simulates the above three visual phenomena observed at a distance and includes three steps: region rescaling, projection (these two steps make up the regular projection process), and differentiated blurring operation.

\textbf{Region Rescaling and Projection.} The 3D point cloud is initially scaled to match the projection cube. The texture and depth are projected onto perpendicular cube planes $Z\in {\N^{W \times H}}$. The side projections are then spliced into multi-perspective images for network input.

% \begin{figure}[htbp]
% \centering
% \includegraphics[width=1\linewidth]{preplusbackbone2.jpg}
% \caption{Perception domain feature extraction. Different perspectives are spliced together into a multi-perspective image. Then a differentiated blurring filter is imposed on the texture projection to simulate the blurring mechanism in the HVS. A modified basic backbone is used to extract the feature vector with an MLP to adjust dimensions.}
% \label{fig:preprocessing}
% \end{figure}

\textbf{Differentiated Blurring Operation.} 
To simulate blurring caused by the viewing distance, we apply a low-pass filter $f(z, \Delta \rho)$ on the texture projection image, which depends on the change in point density $\Delta \rho$.

The low-pass filtering operator can be implemented with a mean filter, averaging the neighborhood $\mathcal{N}(z,R)$ of projected point within a certain radius of $R$ which varies linearly with the change in point density:
\begin{equation}\label{eq13}
\begin{aligned}
    f(z,\Delta \rho) &= \frac{1}{{|N(z,R)|}}\sum_{z_x \in N(z,R)} z_x \\
    \text{s.t.}\quad R &= 
    \begin{cases}
        k\Delta\rho, & \text{if } \Delta\rho > 0 \\
        0, & \text{if } \Delta\rho \leq 0
    \end{cases},
\end{aligned}
\end{equation}
where $\Delta\rho=\tau-\rho$, $\rho$ represents the density of the point cloud, and $\tau$ is an empirical density threshold for the water-tight surface, beyond which density changes no longer affect subjective perception. This blurring operation compensates for density inconsistencies masked by scale reduction.

% The point density $\rho_s$ can be computed by the average point number of local patches $P_s$ located within the corresponding sphere regions $\mathcal{B}(p_s, R_s)$ with the radius $R_s$ \cite{Zhang2021MSGraphSIM}, i.e.
% $\rho_s = \frac{{\left| {{P_s}} \right|}}{{\frac{4}{3}\pi {{{R_s}}^3}}}$, where $|P_s|$ denotes the point number of patch $P_s$. 

\subsubsection{Acquisition of Perception Domain Features}

Following the HVS-based projection, we extract features using ResNet50 from the concatenated image projections and depth projections. This backbone, followed by a 2-layer multilayer perceptron (MLP) with intermediate dimension $d_m$, produces $d$-dimensional features representing immediate visual stimuli. The extracted features are termed the perception domain features, whose extraction operation is denoted as $\phi(\cdot)$.

\subsection{Dataset Scale Normalization}

For anchor feature learning, we assume a uniform distribution of support set samples across impairment scales and assume that participants’ preferences for degradation descriptions of the same sample are distributed within the same quality level. Given the widely acknowledged 5-level quality description for subjective scores \cite{Li2022REQA,Ou2023ICCV}, we employ the BT.500 standard for modeling and categorize the support set into five groups corresponding to the five-grade degradation descriptions \cite{BT500} based on their MOS. For a dataset with the scale $[q_{min},q_{max}]$, the original quality scores ${\dot q}_i$ are initially normalized into the scale of BT.500 using
\begin{align}
{q_i} = \frac{{{\dot q}_i - {q_{\min }}}}{{{q_{\max }} - {q_{\min }}}}*5 + 0.5.
\end{align}
Therefore, the quality level can be obtained by 
\begin{align}
q_{i,L}=round(q_i),
\end{align}
and the confidence degree is denoted as 
\begin{align}
q_{i,R}=\delta(q_i-q_{i,L}),
\end{align}
where $\delta$ signifies the scale factor of confidence degree, set to 1 in this work. Section \ref{sec:confidencescale} delves into the impacts of varying $\delta$ on quality prediction.

\subsection{Anchor Feature Generation Module}

Anchor features are considered the common features for a specific quality, facilitating domain transformation. They are extracted through a weighted combination of sample features with similar degradation, as outlined in \eqref{formu2-13}.

In practice, \eqref{formu2-13} is implemented as a feature distillation operation via an attention transformer network, which takes into count the intra-domain feature interaction and disentangles perception domain features into domain-invariant representations. Then a structured latent space is learned to regularize the disentangled features, defining a structured description domain. From this latent space, anchor features are obtained by aggregating to drive domain transformation.

\subsubsection{Feature Disentanglement}
\label{sec:commonality}

Anchor features rely on a structured distribution within a latent space that reflects the description domain. However, directly constraining the distribution of perception domain features $\phi (x^{(k)})$ is insufficient, as these features represent the dispersed visual stimuli containing both domain-invariant components (encapsulating cross-sample quality-aware commonalities of the same quality level) and sample-specific details (representing unique specificities of sample itself).

\textbf{Domain-invariant Feature Disentanglement.} To comprehensively account for the interaction among samples, we employ an attention transformer \cite{Vaswani2017Transformer} network to amalgamate features. Initially, this module projects $k$ perception domain features of the same quality level in the support set, denoted as ${f_p^{(k)}} = \phi ({x^{(k)}}) \in {\R^{{k} \times {d}}}$, into queries $Q \in {\R^{{k} \times {d_h}}}$, keys $K \in {\R^{{k} \times {d_h}}}$ and values $V \in {\R^{{k} \times {d_h}}}$ using linear projections with the weights $W_q$, $W_k$ and $W_v$. Specifically, $Q= \phi ({x^{(k)}})W_q$, $K= \phi ({x^{(k)}})W_k$, and $V= \phi ({x^{(k)}})W_v$. The cross-sample attention weight is obtained by the scaled dot-product as follows:
\begin{align}\label{eq24}
{\alpha}_A = softmax (\frac{{Q{K^T}}}{{\sqrt {{d}} }}),
\end{align}
which generates the similarity weight based on $K$ and $Q$. Subsequently, the weighted sum of the values $V$ is computed to capture cross-sample commonalities:
\begin{align}\label{eq25}
\begin{split}
Y_{com}&= {\alpha}_AV={\alpha}_A \phi ({x^{(k)}})W_v\approx {\alpha}_A \phi ({x^{(k)}}),
\end{split}
\end{align}
where the extracted domain-invariant features $Y_{com}$ describe cross-sample similarity among features sharing the same quality level, mathematically enabling intra-domain interactions.

To further reduce dependence on sample-specific content, we shuffle mini-batches during training, fostering more robust domain-invariant feature learning.

% where the extracted domain-invariant features $Y_{com}$ measure the cross-sample similarity between $f_p$ and other features sharing the same quality level in the support set. 

\textbf{Module Structure.} This module denoted as $\Phi( \cdot )$ can be formulated as
\begin{align}
\begin{split}
&{y_0} = [ {\phi{({x}_{1}}),\phi{({x}_{2}}), \ldots ,\phi{({x}_{k}})} ],\\
&{Q_i} = {K_i} = {V_i} = FC(d,d_h*n_h)( {{y_{i - 1}}} ),\\
&y_i^{\prime}  = MSA ( {{Q_i},{K_i},{V_i}} ),\\
&{y_i}^{\prime\prime} = FC(d,d_m) ( { {y_i^{\prime} } } ),\\
&{y_i} = FC(d_m,d) ( { {y_i^{\prime\prime} } } ), \quad i = 1, \ldots ,l,\\
&{f_{com}} = [ {f_{{C_1}}},{f_{{C_2}}}, \ldots ,{f_{{C_{K_L}}}} ] = {y_l},
\end{split}
\end{align}
where $f_{com}$ gives the domain-invariant features, which is denoted as $f_{com} = \Phi(\phi(x^{(k)}))$. $y_0$ represents the input $k$ perception domain features of the same quality level. When applying to the support set, $k$ here is marked as $K_L$ in the experimental setup. MSA signifies the multi-head self-attention module~\cite{Vaswani2017Transformer} with the head number of $n_h$ which refers to the feature interaction in \eqref{eq24}. FC means the fully-connected layer~\cite{Vaswani2017Transformer}, which corresponds to the feature mapping in \eqref{eq25}. $d$ denotes the perception domain feature channel, $d_h=d/n_h$ represents the head dimension for applying MSA, and $d_m$ signifies the intermediate dimensions for cascaded fully-connected layers. $l$ is the number of layers. 

% The described pipeline outlines the feature processing for support set samples. For subsequent relevance measurements, the query set sample follows the same pipeline to reduce feature space bias. Here, the feature interaction (MSA) is inactive, and $\Phi( \cdot )$ functions as an MLP.

Note that the aforementioned pipeline depicts the feature process for support set samples. In the subsequent relevance measurements, the single sample in the query set undergoes the same pipeline to mitigate feature space bias. In this case, the feature interaction operation (MSA) does not take effect, and $\Phi( \cdot )$ is equivalent to an MLP.

\subsubsection{Description Domain Establishment}

Samples with the same quality level share degradation descriptions, suggesting that intra-domain relevance exists and disentangled features for each level can be structured within a latent space. This structured space is trained with constraints on the support set to reproduce the function of the description domain.

\textbf{Distribution Regularization.} Features from the same quality level are expected to cluster within the latent space and vice versa. To achieve this, we apply a weighted InfoNCE \cite{Oord2018InfoNCE} loss. Extracted domain-invariant features are L2-normalized as $f^{(k)}_{n}={L_2norm}(\Phi(\phi ({x^{(k)}_{n}})))$. Positive pairs consist of non-overlapping samples with the same quality level, while negative pairs consist of non-overlapping samples with different quality levels.

After defining the similarity between two features as
\begin{align}
h( {{{f}_{n}},{{f}_{m}}} ) = \exp ( {\frac{{{{f}_{n}} \cdot {{f}_{m}}}}{{{{\left\| {{{f}_{n}}} \right\|}_2} \cdot {{\left\| {{{f}_{m}}} \right\|}_2}}} \cdot \frac{1}{\tau }} ),
\end{align}
where $\tau$ controls the range of the similarity results. The distribution regularization function for a positive pair of feature ${{f_{n,k}}}$ is formulated as
\begin{align}
\mathcal{L}_{dis}^{n,k} = -\log [ {\frac{{{w_{pos}}h( {{f_{n,k}},f{'_{n,k}}} )}}{{{w_{pos}} h( {{f_{n,k}},f{'_{n,k}}} ) + \sum\limits_{m,j,m \ne n} {{w_{neg}}h} ( {{f_{n,k}},f{'_{m,j}}} )}}} ],
\end{align}
where ${{w_{pos}}}$ for positive pairs $[{f_{n,k}},{f'_{n,k}}]$ and ${{w_{neg}}}$ for negative pairs $[{{f_{n,k}},f{'_{m,j}}}]$ denote the later-defined probability weights.

To further promote the quality-aware structured distributions, the probability weights ${{w_{pos}}}$ and ${{w_{neg}}}$ regarding quality variations are applied. Specifically, ${{w_{pos}}}$ measures the probability of $[{f_{n,k}},{f'_{n,k}}]$ forming positive pairs based on their quality difference, while ${{w_{neg}}}$ gauges the probability of $[{{f_{n,k}},f{'_{m,j}}}]$ constituting negative pairs based on their quality difference, i.e.
\begin{align}
{w_{pos}} = \frac{1}{{{{( {q_{n,k}^{} - q{'_{n,k}}^{}} )}^2} + 1}},
\end{align}
and
\begin{align}
{w_{neg}} = 1 - \frac{1}{{{{( {q_{n,k}^{} - q{'_{m,j}}^{}} )}^2} + 1}},
\end{align}
where $q_{n,k}$ denotes the MOS for the feature $f_{n,k}$. 

For a mini-batch, the final distribution loss function to regularize the feature distribution for the domain-invariant features is defined as follows:
\begin{align}
\begin{split}
{\mathcal{L}_{dis}} = \frac{1}{{K_L\times N}}\sum\limits_{k = 1}^{K_L}\sum\limits_{n = 1}^{N} {\mathcal{L}_{dis}^{n,k}}  (Y_{com}) ,
\end{split}
\end{align}
where $K_L$ signifies the number of samples with the same quality level in a mini-batch, and $N$ denotes the quality level number.

\subsubsection{Description Domain Feature Aggregation}
\label{sec:hallucinated}

Once a structured latent space is learned, cluster centers characterize the description domain for each quality level, serving as anchor features for quality distinction in the query set. These anchor features provide auxiliary information beyond the perception domain, thereby helping sample features achieve domain transformation more easily. To produce anchor features, domain-invariant features from samples of the same quality level are aggregated. The aggregation denoted as $F(\cdot)$ can be formulated as
\begin{align}\label{eq37}
\begin{split}
{f_R} = \frac{1}{K_L}\sum\limits_{k=1}^{K_L} {\Phi (\phi ({x^{(k)}}))} \approx \underbrace{\frac{1}{K_L}e\alpha _A}_{W_{\alpha}} \phi ({x^{(k)}}),
\end{split}
\end{align}
where $f_R$ is the obtained anchor feature for a specific quality level. $e$ is the unit vector.

\subsection{Relevance Mapping Module for Stage-1}
\label{sec:mapping1}

\subsubsection{Inter-domain Relevance Mapping}

Referring to \eqref{formu2-14}, the inter-domain relevance between the testing sample feature and five anchor features is measured and then mapped into the quality level. However, the mapping operation in \eqref{formu2-14}, denoted as $F_1(A, B) =(A \odot B)\times W_1$ where $\odot$ represents the element-wise multiplication, limits the information usage and may lead to gradient truncation due to the interdependence of gradients under the Leibniz product rule. This operation is thus not ideal for neural network design.

To address this, we concatenate the testing sample features with anchor features, enriching the testing features with description domain information. A mapping network then measures inter-domain relevance and maps it into quality-level probability scores, denoted as $F_2(A, B) =[A ; B]\times W_2$. The final quality level is determined by the maximum probability score.

\textbf{Module Structure.} This module denoted as $G( \cdot )$ can be formulated as
\begin{align}
\begin{split}
{g_{i,j,1}} &= \Phi_{x_i} \oplus {\Phi_{R,j}}, j = 1, \ldots ,5,\\
{g_{i,j,2}} &= FC(d\times 2, d_m)({g_{i,j,1}}),\\
{g_{i,j,3}} &= Relu({g_{i,j,2}}),\\
{g_{i,j,4}} &= FC(d_m, 1)({g_{i,j,3}}),\\
{g_{i,l}} &= argmax(Softmax([g_{i,j,4},j = 1, \ldots ,5])),
\end{split}
\end{align}
where $g_{i,l}$ is the output of $G$ identified as the predicted quality level of the testing sample $x_i$, $\Phi_{x_i}=\Phi(\phi (x_i))$ in which the sample $x_i$ from the query set is mapped into the learned latent space to prevent additional errors due to offsets in the feature space, $\Phi_{R,j}$ gives the anchor feature for the $j$-th quality level, and $\oplus$ represents the concatenation operation. FC means the fully-connected layers which are characterized by the input and output channel number. $d$ denotes the perception domain feature channel, and $d_m$ signifies the intermediate dimensions for cascaded fully-connected layers.

\subsubsection{Boundary Regularization} 

To enhance feature boundary learning across quality levels in the latent space, a cross entropy loss is applied:

\small
\begin{align}
{L_{cls}} = \frac{1}{{K_Q}}\sum\limits_{i = 1}^{K_Q} {\sum\limits_{j = 1}^N  -  } {q_{i,j,L}}\log {{\hat q}_{i,j,L}} - ( {1 - {q_{i,j,L}}} )\log ( {1 - {{\hat q}_{i,j,L}}} ),
\end{align}
where ${\hat q}_{i,j,L}$ gives the predicted probability for each quality level, $q_{i,j,L}$ represents the ground truth, $K_Q$ is the number of samples in the query set of a mini-batch, and $N$ denotes the quality level number.

Boundary regularization links the clustering centers in the learned latent space with the optimization objectives, \textit{i.e.,} the quality levels. This promotes the development of distinctions among features with different quality levels in the learned latent space.

\subsection{Relevance Mapping Module for Stage-2}
\label{sec:stage2}

\subsubsection{Intra-domain Relevance Mapping}
\label{sec:mapping2}

Referring to \eqref{formu2-15}, the intra-domain relevance in Stage-2 quantifies the relevance between the testing sample in the query set and support set samples within the local description domain, yielding a confidence degree.

Once the quality level is determined, the decision space is constrained to a local area within the description domain corresponding to a specific quality level. Then similar to the operation in the previous subsection, we measure and map relevance by concatenating features of the testing sample and neighboring support samples to predict their confidence biases, optimizing the usage of support set annotations. The testing sample’s confidence degree is then calculated from the predicted confidence biases and the confidence degrees of the neighboring support set samples.

\textbf{Module Structure.} When the $j$-th quality level is determined in Stage-1, this module denoted as $H( \cdot )$ is formulated as
\begin{align}
\label{formu19}
\begin{split}
{h_{i,k,1}}& = {\Phi _{x_i}} \oplus {\Phi _{x_{k,j}}},\quad k = 1, \cdots ,K_L,\\
{h_{i,k,2}}& = FC(d \times 2, d_m)({h_{i,k,1}}),\\
{h_{i,k,3}}& = Relu({h_{i,k,2}}),\\
{h_{i,k,4}}& = FC(d_m, 1)({h_{i,k,3}}),\\
{h_{i,k}}& = Sigmoid({h_{i,k,4}})-0.5,\\
{h_{i,r}}& = mean([q_{1,j,R}+{h_{i,1}},q_{2,j,R}+{h_{i,2}}, \cdots ,q_{K_L,j,R}+{h_{{i,K_L}}}]),
\end{split}
\end{align}
where $h_{i,k}$ is the predicted confidence biases between the testing sample $x_i$ in the query set and the $k$-th support set sample, derived from the relevance measurements and confined within the confidence scale of $[-0.5,0.5]$. $h_{i,r}$ gives the output of $H$ identified as the predicted confidence degree of the testing sample $x_i$. $\Phi _{x_i}=\Phi(\phi (x_i))$ represents the disentangled description domain features of the sample $x_i$ in the query set, and $\Phi _{x_{k,j}} = \Phi(\phi (x_{k,j}))$ indicates the disentangled feature of $k$-th sample identified as the $j$-th quality level by Stage-1 in the support set. $q_{k,j,R}$ represents the confidence degree of the $k$-th support set sample with the $j$-th quality level, and $K_L$ denotes the sample number for the $j$-th quality level in the support set of a mini-batch.

\subsubsection{Quality Combination Module}
\label{sec:combination}

Given the support set samples $x^{(k)}$, the continuous quality score of the sample $x_i$ in the query set can be obtained by
\begin{align}
\begin{split}
{\hat q}_i&={\hat q}_{i,L}+{\hat q}_{i,R}\\
&={G(F(\Phi(\phi(x^{(k)}))), \Phi(\phi(x_i)))} + {H(\Phi(\phi(x^{(k)})),\Phi(\phi(x_i)))},
\end{split}
\end{align}
where ${\hat q}_i$ is the desired predicted continuous quality score.

\subsubsection{Quality-aware Regularization}

To ensure the quality sensitivity of  feature commonalities and achieve final quality prediction, quality-aware regularization is applied using the Pearson linear correlation coefficient (PLCC) and Spearman rank-order correlation coefficient (SROCC) as loss functions. Among them, the SROCC loss can be computed from PLCC loss using
\begin{align}\label{eq:rankloss}
\begin{split}
loss_{srocc}( {{q^{(K_Q)}}, {{\hat q}^{(K_Q)}}} ) 
&=los{s_{plcc}}( {P( {{q^{(K_Q)}}} ), P( {{{\hat q}^{(K_Q)}}} )} )\\
&=PLCC ({P( {{q^{(K_Q)}}} ), P( {{{\hat q}^{(K_Q)}}} )} ),
\end{split}
\end{align}
where ${\hat q}^{(K_Q)}$ gives the predicted quality scores of the query set samples in a mini-batch, and $q^{(K_Q)}$ indicates the ground truth MOS. $P$ represents the rank function calculated by the Heaviside step function which is not differentiable and therefore approximated by a constrained linear program~\cite{Blondel2020Fastsorting, liu2024VCIP}, denoted as
\begin{align}
P_{\varepsilon Q}({x})=P_{\varepsilon Q}(-{x}, {\rho})=P_{Q}(-{x} / \varepsilon, {\rho})
\end{align}
\begin{align}
P_Q({z}, {w})=\underset{{\mu} \in \mathcal{P}({w})}{\operatorname{argmax}}\langle{z}, {\mu}\rangle-Q({\mu})=\underset{{\mu} \in \mathcal{P}({w})}{\operatorname{argmin}} \frac{1}{2}\|{\mu}-{z}\|^2
\end{align}
where $\rho=(n, n-1, \ldots, 1)$. $\underset{{\mu} \in \mathcal{P}({w})}{\operatorname{argmax}}\langle{z}, {\mu}\rangle$ is the linear program, and $Q({\mu})=\frac{1}{2}\|{\mu}\|^2$ is quadratic regularization. $\mathcal{P}({w})=\operatorname{conv}(\left\{{w}_\sigma: \sigma \in \Sigma\right\}) \subset \mathbb{R}^n$ represents the convex hull of permutations of $w$.

As a result, the quality-aware regularization is give by
\begin{align}
\mathcal{L}_{reg} = - loss_{plcc}( {{q^{(k)}},{{\hat q}^{(k)}}} ) - loss_{srocc}( {{q^{(k)}},{{\hat q}^{(k)}}} ).
\end{align}

\subsection{Overall Loss}

The overall training loss is the weighted sum of distribution loss, boundary loss and quality-aware loss, which leads to
\begin{align}
\mathcal{L} = \lambda_1\mathcal{L}_{dis} + \lambda_2\mathcal{L}_{cls} + \lambda_3\mathcal{L}_{reg},
\end{align}
where $\lambda_1$, $\lambda_2$ and $\lambda_3$ are the weighting factors.

\begin{figure*}[t]
\centering
\includegraphics[width=1\linewidth]{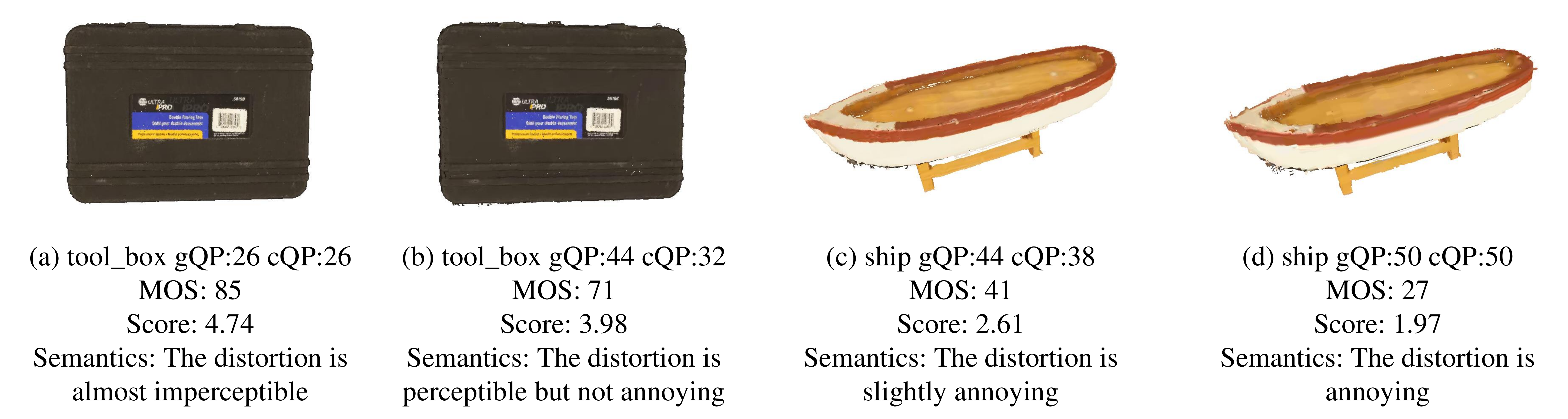}
\caption{Exemplary point clouds with the subjective MOS, the predicted quality score of the proposed D$^3$-PCQA and the degradation description corresponding to the predicted quality level on the WPC dataset. The predicted quality score and the degradation description align with subjective perception.}
\label{fig:examples}
\end{figure*}

\section{Experimental Results and Analyses}
\label{sec:experiments}

\subsection{Implementation Details}

To evaluate the performance of the proposed D$^3$-PCQA, we conduct evaluation experiments on different datasets, \textit{i.e.,} SJTU-PCQA~\cite{Yang2020TMM3DTO2D} and WPC datasets~\cite{Su2019WPC,Liu2022WPC}. 

To ensure fairness and minimize content bias, following the practice in \cite{shan2022GPANet}, a 5-fold cross-validation strategy is adopted for the SJTU-PCQA and WPC datasets due to the limited dataset scale. Specifically, for SJTU-PCQA, the dataset is split into the training-testing sets with the ratio of 7:2 according to the reference point clouds for each fold, and the average performance in the testing sets is recorded. Similarly, for WPC, the training-testing ratio is set to 4:1. Notably, there is no content overlap among the training and testing sets. During training, the training set is evenly divided into the support set and query set based on the reference point clouds. 

To ensure consistency between predicted scores and MOSs, a nonlinear Logistic-4 regression is applied to map the predicted scores to the same dynamic range, following the recommendations suggested by the Video Quality Experts Group (VQEG) \cite{Antkowiak2000vqeg,2006logistic5}. The employed evaluation metrics include PLCC and root mean square error (RMSE) for prediction accuracy, SROCC and Kendall rank-order correlation coefficient (KROCC) for prediction monotonicity. Larger PLCC, SROCC, or KROCC values indicate better performance, while a smaller RMSE reflects better performance.

The Adam optimizer is employed with a weight decay of 1e-4, and an initial learning rate of 5e-5. Within each batch, the sample number for the support set is set to 30, with the sample number $K_L$ assigned to each quality level set to 6, and the sample number $K_Q$ for the query set or the testing set is set to 15. The default training duration for the model is 300 epochs. The factor $\delta$ to control the scale of confidence degree is set to 1. The sizes $W$ and $H$ of projection planes are both 224. The dimension $d$ of extracted perception domain features is 1024. The intermediate dimension $d_m$ of the cascaded fully-connected layers in both Stage-1 and Stage-2 is also 1024. The layer number $l$ in the feature disentanglement module is 3, and the head number $n_h$ is set to 8. % The weighting factors $\lambda_1$, $\lambda_2$ and $\lambda_3$ are set to 1.

% To augment the training sets, for projection-based methods, each point cloud is projected to 12 versions of images by rotating it to 12 different viewpoints. The 12 viewpoints are uniformly placed around the object based on the 12 polyhedron vertices of a regular icosahedron \cite{Alexiou2019projection,Liu2021PQANet}. For 3D-based methods, random rotation within the range of $[{0^\circ },{360^\circ })$ is invoked during training. 

\subsection{Overall Performance}
\label{sec:overall}

We compare the performance of the proposed D$^3$-PCQA with prevalent 3D-based NR methods, including ResSCNN~\cite{Liu2022LSPCQA}, and GPA-Net~\cite{shan2022GPANet}, and prevalent projection-based NR methods, including PQA-Net~\cite{Liu2021PQANet} and IT-PCQA~\cite{Yang2021ITPCQA}. The results are summarized in Table \ref{tab:overall}, with the best results highlighted in bold and the second-best results highlighted with underline.

% Table generated by Excel2LaTeX from sheet '整体4'
\begin{table*}[htbp]
  \centering
  \caption{Overall performance. The best results are highlighted in bold, while the second-best results are underlined. Modal P means that the methods operate on the raw 3D data, and Modal I indicates that the methods operate on the point cloud projection.}
   \begin{tabular}{l|c|cccc|cccc}
    \hline
    \multirow{2}[4]{*}{Method} & \multirow{2}[4]{*}{Modal} & \multicolumn{4}{c|}{SJTU-PCQA}  & \multicolumn{4}{c}{WPC} \\
\cline{3-10} &       & PLCC  & SROCC & KROCC & RMSE  & PLCC  & SROCC & KROCC & RMSE \\
    \hline
    ResSCNN  & P     & \underline{0.867}  & \underline{0.858}  & \underline{0.654}  & 1.113  & \underline{0.747}  & \underline{0.728}  & \underline{0.517}  & 16.131  \\
          GPA-Net & P     & 0.864  & 0.855  & 0.632  & \underline{1.058}  & 0.722  & 0.704  & 0.497  & 15.784  \\
 PQA-Net & I     & 0.859  & 0.836  & 0.646  & 1.072  & 0.718  & 0.703  & 0.508  & \underline{15.073}  \\
          IT-PCQA & I     & 0.609  & 0.597  & 0.381  & 1.976  & 0.570  & 0.552  & 0.416  & 17.126  \\
          D$^3$-PCQA (ours) & I     & \textbf{0.940} & \textbf{0.942} & \textbf{0.798} & \textbf{0.792} & \textbf{0.855} & \textbf{0.862} & \textbf{0.669} & \textbf{11.800} \\
    \hline
\end{tabular}
  \label{tab:overall}%
\end{table*}%

We can see from Table \ref{tab:overall} that: i) the proposed D$^3$-PCQA exhibits robust and outstanding performance across different datasets; ii) the proposed D$^3$-PCQA promotes the easier domain transformation, which is more evident on datasets with huge domain discrepancy; iii) D$^3$-PCQA consumes a certain degree of precise fitting ability to individual sample in exchange for fitting intra- and inter-domain relevance. Compared to the reduction resulting from the former, the improvement brought about by the latter is notably more pronounced in quality prediction, especially on complex datasets.

Additionally, we show some examples of distorted point clouds with the subjective MOS, the predicted quality score of the proposed D$^3$-PCQA and the degradation description corresponding to the predicted quality level in Fig. \ref{fig:examples}. It can be observed that the predicted quality score and the degradation description align with subjective perception.

\subsection{Generalization Performance}

In practical scenarios, the generalization ability assumes heightened significance. In this subsection, we evaluate the generalization ability of the proposed D$^3$-PCQA through the cross-dataset experiment. Specifically, the proposed D$^3$-PCQA and other NR-PCQA methods are trained on the SJTU-PCQA dataset and tested on the WPC dataset. Then the two datasets are switched and the experiment is repeated. The cross-dataset evaluation results are shown in Table \ref{tab:crosstest1}. The best results are highlighted in bold, and the second-best results are highlighted with underline.

% Table generated by Excel2LaTeX from sheet '跨数据集3'
\begin{table*}[htbp]
  \centering
  \caption{Generalization performance of NR-PCQA methods. The best results are highlighted in bold, while the second-best results are underlined. The proposed D$^3$-PCQA exhibits the highest generalization ability under all testing conditions.}
    \begin{tabular}{l|cccc|cccc}
    \hline
          & \multicolumn{4}{c|}{SJTUPCQA-WPC} & \multicolumn{4}{c}{WPC-SJTUPCQA} \\
\cline{2-9}          & PLCC  & SROCC & KROCC & RMSE  & PLCC  & SROCC & KROCC & RMSE \\
    \hline
    ResSCNN & 0.267  & 0.258  & 0.198  & \underline{20.158}  & \underline{0.570}  & \underline{0.536}  & 0.410  & \underline{2.018}  \\
    GPA-Net & \underline{0.430}  & \underline{0.419}  & \underline{0.227}  & 20.267  & 0.537  & 0.528  & \underline{0.457}  & 2.165  \\
    PQA-Net & 0.285  & 0.283  & 0.199  & 20.432  & 0.489  & 0.478  & 0.333  & 2.232  \\
    IT-PCQA & 0.296  & 0.288  & 0.177  & 21.482  & 0.333  & 0.325  & 0.226  & 2.337  \\
    D$^3$-PCQA & \textbf{0.485} & \textbf{0.434} & \textbf{0.297} & \textbf{20.052} & \textbf{0.751} & \textbf{0.767} & \textbf{0.568} & \textbf{1.604} \\
    \hline
\end{tabular}
  \label{tab:crosstest1}%
\end{table*}%

% $\uparrow$

We can see from Table \ref{tab:crosstest1} that: i) the proposed D$^3$-PCQA exhibits superior generalization ability compared with other NR-PCQA methods under all testing conditions, highlighting its effectiveness; ii) the utilization of the training data significantly affects the generalization performance of NR-PCQA models. In general, 3D-based NR-PCQA methods outperform projection-based methods, which is explained by the high efficiency in utilizing the training data by 3D-based backbones. However, the performance of the proposed D$^3$-PCQA, despite using a only projection-based backbone, still demonstrates superior generalization performance compared to other learning-based NR-PCQA methods, demonstrating the role of the proposed domain transformation theory; iii) the coverage of the training data also affects the generalization performance of NR-PCQA methods. Methods including the proposed D$^3$-PCQA trained on the complex datasets benefit from training data with broader coverage, leading to higher generalization performance; iv) combining Table \ref{tab:overall} and Table \ref{tab:crosstest1}, it can be seen that the proposed D$^3$-PCQA does not show significant performance gains when training on simple datasets, as it relies on exploring relevance between samples.

\subsection{Effect of Loss Functions}

\begin{figure}[t]
\centering
\includegraphics[width=0.8\linewidth]{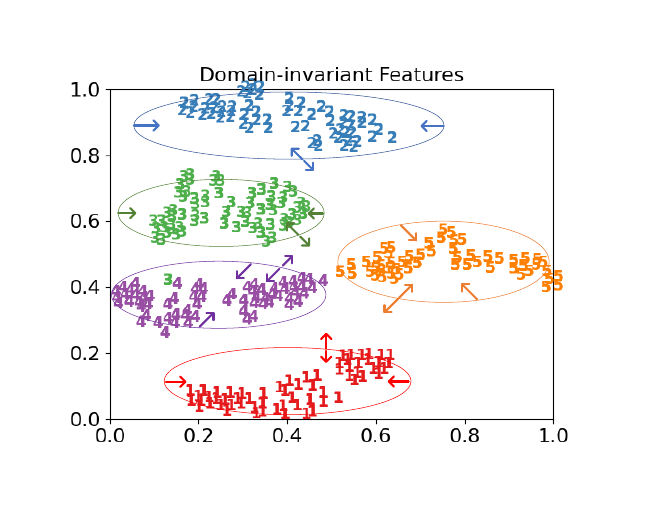}
\caption{T-SNE plot of disentangled features on the WPC dataset. Each number indicates a feature of the corresponding quality level. For the disentangled domain-invariant features, through distribution regularization, features of the same quality level are clustered, and features of different quality levels tend to show distinct boundaries, forming a structured but not hierarchical enough transitional feature space.}
  \label{fig:cluster}
\end{figure}

The proposed network is trained using the distribution loss, boundary loss and quality-aware loss. In this subsection, we evaluate the effect of each loss function on the WPC dataset. We use $\mathcal{L}_{cls}$ to train the network as a benchmark. To demonstrate the effectiveness of establishing the description domain, $\mathcal{L}_{cls} + \mathcal{L}_{dis}$ (whole Stage-1) is used as the loss function to repeat the trial. Note that when $\mathcal{L}_{reg}$ is not used, the final predicted quality score is only $\hat q = {\hat q}_L$ with ${\hat q}_R$ forced to 0. Besides, to visualize the effect of the distribution regularization, we give the t-SNE plots of the disentangled features on the WPC dataset as shown in Fig. \ref{fig:cluster}. Finally, $\mathcal{L}_{cls} + \mathcal{L}_{dis} + \mathcal{L}_{reg}$ (add Stage-2) is used as the loss function to train the network, which achieves the final performance of the proposed model. The performance with different loss functions are shown in Table \ref{tab:ablation}.

% Under the latter two conditions, the change of $\mathcal{L}_{cls}$ during training is shown in Fig. \ref{fig:training} to further demonstrate the promoting effect of Stage-2 on Stage-1. 

% \begin{figure}[htbp]
% \centering
% \includegraphics[width=0.8\linewidth]{training.pdf}
% \caption{\RE{$\mathcal{L}_{cls}$ during training with different loss functions. This shows that the involvement of Stage-2 can promote the training of Stage-1.}}
% \label{fig:training}
% \end{figure}

% 两个表格拼一起
\begin{table}[htbp]
  \centering
  \caption{Performance with different loss functions.}
    \begin{tabular}{ccc|cccc}
    \hline
    \multicolumn{1}{l}{$\mathcal{L}_{cls}$} & \multicolumn{1}{l}{$\mathcal{L}_{dis}$} & \multicolumn{1}{l|}{$\mathcal{L}_{reg}$} & \multicolumn{1}{l}{PLCC} & \multicolumn{1}{l}{SROCC} & \multicolumn{1}{l}{KROCC} & \multicolumn{1}{l}{RMSE} \\
    \hline
    \Checkmark     & \XSolidBrush     & \XSolidBrush     &   0.633    &  0.608     &  0.499     & 17.170 \\
    \Checkmark     & \Checkmark     & \XSolidBrush     & 0.677      & 0.690      & 0.567      & 16.668 \\
    \Checkmark     & \Checkmark     & \Checkmark     & \textbf{0.855} & \textbf{0.862} & \textbf{0.669} & \textbf{11.800} \\
    \hline
    \end{tabular}%
  \label{tab:ablation}
\end{table}

 % \XSolidBrush     & \XSolidBrush     & \Checkmark     & 0.723  & 0.714  & 0.531  & 15.575 \\

We can see from Table \ref{tab:ablation} that: i) the sole use of $\mathcal{L}_{cls}$ leads to unsatisfactory performance; ii) $\mathcal{L}_{dis}$ can promote domain transformation by exploiting intra-domain relevance to reproduce the function of the description domain. Particularly, Fig. \ref{fig:cluster} demonstrates that the distribution regularization forces the aggregation within the same quality level and the distinct boundaries between different quality levels for the disentangled features; iii) $\mathcal{L}_{reg}$ leads to obvious gain in performance, which demonstrates that $\mathcal{L}_{reg}$ can better promote the learning of quality-aware features and facilitate the fine-grained quality judgment.

% , and the proposed cross-stage training mechanism is effective for improving the performance of the model.

\subsection{Effect of Projection Strategy}

In this work, we propose the HVS-based projection to address the issue of information loss caused by the projection operation. In this subsection, we conduct evaluation experiments on the WPC dataset to demonstrate the effectiveness of the proposed method. Specifically, the network without the HVS-based projection is trained and tested under the same conditions as the original network. The performance comparison is shown in Table \ref{tab:projection}. 

\begin{table}[htbp]
  \centering
  \caption{Performance improvement using the proposed HVS-based projection on the WPC dataset.}
    \begin{tabular}{l|cccc}
    \hline
          & \multicolumn{1}{l}{PLCC} & \multicolumn{1}{l}{SROCC} & \multicolumn{1}{l}{KROCC} & \multicolumn{1}{l}{RMSE} \\
    \hline
    W/o HVS projection & 0.839      & 0.838 & 0.636 & 12.380\\
    Proposed network & \textbf{0.855} & \textbf{0.862} & \textbf{0.669} & \textbf{11.800}\\
    \hline
    \end{tabular}%
  \label{tab:projection}%
\end{table}%

We can see from Table \ref{tab:projection} that the proposed HVS-based projection to handle information loss exhibits performance improvements. It enhances the utilization of training samples, and allows the model to respond to the masked distortions from a different perspective, thus improving the quality prediction capability of the model.

\subsection{Effect of Feature Disentanglement}

In this subsection, we conduct evaluation experiments on the WPC dataset to demonstrate the effectiveness of the feature disentanglement operation. Specifically, the network without the feature disentanglement is trained and tested under the same conditions as the original network, in which case the distribution regularization is directly imposed on the extracted perception domain features. The results are shown in Table \ref{tab:disentanglement}.

% Table generated by Excel2LaTeX from sheet '提高投影性能'
\begin{table}[htbp]
  \centering
  \caption{Overall performance on the WPC dataset of whether to adopt the feature disentanglement.}
    \begin{tabular}{l|cccc}
    \hline
          & \multicolumn{1}{l}{PLCC} & \multicolumn{1}{l}{SROCC} & \multicolumn{1}{l}{KROCC} & \multicolumn{1}{l}{RMSE} \\
    \hline
    W/o disentanglement & 0.748      & 0.742 & 0.555 & 15.102\\
    Proposed network & \textbf{0.855} & \textbf{0.862} & \textbf{0.669} & \textbf{11.800}\\
    \hline
    \end{tabular}%
  \label{tab:disentanglement}%
\end{table}%

We can see from Table \ref{tab:disentanglement} that directly constraining the distribution of perception domain features diminishes the performance of the network. This outcome arises from the failure to alleviate the effect of sample-specific components within the perception domain features, leading to the difficulty in domain transformation.

\subsection{Effect of Basic Backbones}

In this subsection, we evaluate the performance of the proposed model with different basic backbones during extracting the perception domain features. Specifically, different basic backbones, including AlexNet, VGG16, ResNet34 and ResNet50, are used in the feature extraction module, and the performance of the model is evaluated on the WPC dataset as shown in Table \ref{tab:backbones}.

% Table generated by Excel2LaTeX from sheet 'tau的选择'
\begin{table}[htbp]
  \centering
  \caption{Overall performance on the WPC dataset with different backbones.}
    \begin{tabular}{c|cccc}
    \hline
        & \multicolumn{1}{l}{PLCC} & \multicolumn{1}{l}{SROCC} & \multicolumn{1}{l}{KROCC} & \multicolumn{1}{l}{RMSE} \\
    \hline
    \multicolumn{1}{c|}{AlexNet} & 0.819  & 0.782  & 0.569     & 13.068 \\
    \multicolumn{1}{c|}{VGG16} & 0.813 & 0.804 & 0.612 & 13.090 \\
    \multicolumn{1}{c|}{ResNet34} & 0.805 & 0.789 & 0.581 & 13.498 \\
    \multicolumn{1}{c|}{ResNet50} & \textbf{0.855} & \textbf{0.862} & \textbf{0.669} & \textbf{11.800} \\
    \hline
    \end{tabular}%
  \label{tab:backbones}%
\end{table}%

We can see from Table \ref{tab:backbones} that the proposed method exhibits good performance with multiple common backbones, highlighting its superiority and generalization. Among these common backbones, ResNet50 as the basic backbone generates the best performance due to its stronger fitting ability brought by more layers.

\subsection{Effect of Mapping functions}

In this subsection, we evaluate the performance of applying different mapping functions, including $F_1(A, B) =(A \odot B)\times W_1$ and $F_2(A, B) =[A ; B]\times W_2$. Specifically, the two mapping functions are used for relevance mapping in the proposed network respectively. Meanwhile, the input channels of the following cascaded fully-connected layers are adjusted accordingly. The performance on the WPC dataset is shown in Table \ref{tab:mapping}.

% Table generated by Excel2LaTeX from sheet 'tau的选择'
\begin{table}[htbp]
  \centering
  \caption{Overall performance on the WPC dataset with different mapping functions.}
    \begin{tabular}{c|cccc}
    \hline
        & \multicolumn{1}{l}{PLCC} & \multicolumn{1}{l}{SROCC} & \multicolumn{1}{l}{KROCC} & \multicolumn{1}{l}{RMSE} \\
    \hline
    \multicolumn{1}{c|}{$F_1(A, B)$} & 0.817 & 0.821 & 0.620 & 12.960 \\
    \multicolumn{1}{c|}{$F_2(A, B)$} & \textbf{0.855} & \textbf{0.862} & \textbf{0.669} & \textbf{11.800} \\
    \hline
    \end{tabular}%
  \label{tab:mapping}%
\end{table}%

We can see from Table \ref{tab:mapping} that $F_2(A, B)$ adopted as the mapping function generates better performance due to the potential drawbacks of $F_1(A, B)$ in information utilization and gradient propagation.

\subsection{Effect of Confidence Degree Scale}
\label{sec:confidencescale}

In this subsection, we conduct evaluation experiments to demonstrate the effect of different confidence degree scales on quality prediction. Specifically, the performance of the model using different confidence scales which are controlled by $\delta$ is evaluated on the WPC dataset, as shown in Table \ref{tab:confidencescale}.

% Table generated by Excel2LaTeX from sheet 'tau的选择'
\begin{table}[htbp]
  \centering
  \caption{Overall performance on the WPC dataset with different confidence scale factors $\delta$.}
    \begin{tabular}{c|cccc}
    \hline
    $\delta$    & \multicolumn{1}{l}{PLCC} & \multicolumn{1}{l}{SROCC} & \multicolumn{1}{l}{KROCC} & \multicolumn{1}{l}{RMSE} \\
    \hline
    \multicolumn{1}{c|}{0.5} & 0.846  & 0.838  & 0.636     & 12.108 \\
    \multicolumn{1}{c|}{1} & \textbf{0.855} & \textbf{0.862} & \textbf{0.669} & \textbf{11.800} \\
    \multicolumn{1}{c|}{2} & 0.848  & 0.850  & 0.662     & 12.032 \\
    unlimited & 0.807  & 0.811  & 0.612     & 13.320 \\
    \hline
    \end{tabular}%
  \label{tab:confidencescale}%
\end{table}%

We can see from Table \ref{tab:confidencescale} that: i) an unlimited scale is detrimental to quality prediction performance, in which the proposed model degenerates into the conventional regression-based model, and there will be no differences compared to the existing implementations; ii) a confidence scale of $\delta=1$ with the best performance is determined which is also in line with intuition and experience. 

\subsection{Effect of Probability Weights in Distribution Regularization}

In this subsection, we evaluate the effect of probability weights $w_{pos}$, $w_{neg}$ in the distribution regularization function. Specifically, the performance of the model without the probability weights $w_{pos}$, $w_{neg}$ is evaluated as shown in Table \ref{tab:weightposandneg}.

% Table generated by Excel2LaTeX from sheet 'tau的选择'
\begin{table}[htbp]
  \centering
  \caption{Overall performance on the WPC dataset of whether to apply the probability weights $w_{pos}$ and $w_{neg}$.}
    \begin{tabular}{cc|cccc}
    \hline
    \multicolumn{1}{l}{$w_{pos}$} & \multicolumn{1}{l|}{$w_{neg}$} & \multicolumn{1}{l}{PLCC} & \multicolumn{1}{l}{SROCC} & \multicolumn{1}{l}{KROCC} & \multicolumn{1}{l}{RMSE} \\
    \hline
    \XSolidBrush     & \Checkmark     & 0.839  & 0.834  & 0.642  & 12.300  \\
    \Checkmark     & \XSolidBrush     & 0.825  & 0.835  & 0.635  & 12.866  \\
    \Checkmark     & \Checkmark     & \textbf{0.855}  & \textbf{0.862}  & \textbf{0.669}  & \textbf{11.800}  \\
    \hline
    \end{tabular}%
  \label{tab:weightposandneg}%
\end{table}%

We can see from Table \ref{tab:weightposandneg} that the probability weights in the distribution regularization exhibit a gain in performance by further regularizing the feature distribution in the description domain.

\section{Conclusion}
\label{sec:conclusion}

In this study, we propose a novel NR-PCQA method called D$^3$-PCQA, which considers quality assessment as a domain transformation from the perception domain to the quality domain. To reduce domain discrepancy,  we establish a new intermediate domain, namely the description domain, by exploiting domain relevance and learning a structured latent space. The anchor features derived from the learned structured latent space are generated as cross-domain auxiliary information to promote domain transformation. Furthermore, the established description domain decomposes quality prediction into the degradation description prediction and the confidence degree prediction, providing a semantic explanation for the predicted quality scores. Experimental results demonstrate the effectiveness of D$^3$-PCQA, which achieves robust overall performance and outstanding generalization compared with existing NR-PCQA methods.

\ifCLASSOPTIONcaptionsoff
  \newpage
\fi

%\cite{}

\bibliography{manuscript}
\bibliographystyle{IEEEtran}
% \vspace{-10 mm}

\begin{IEEEbiography}
[{\includegraphics[width=1in,height=1.25in,clip,keepaspectratio]{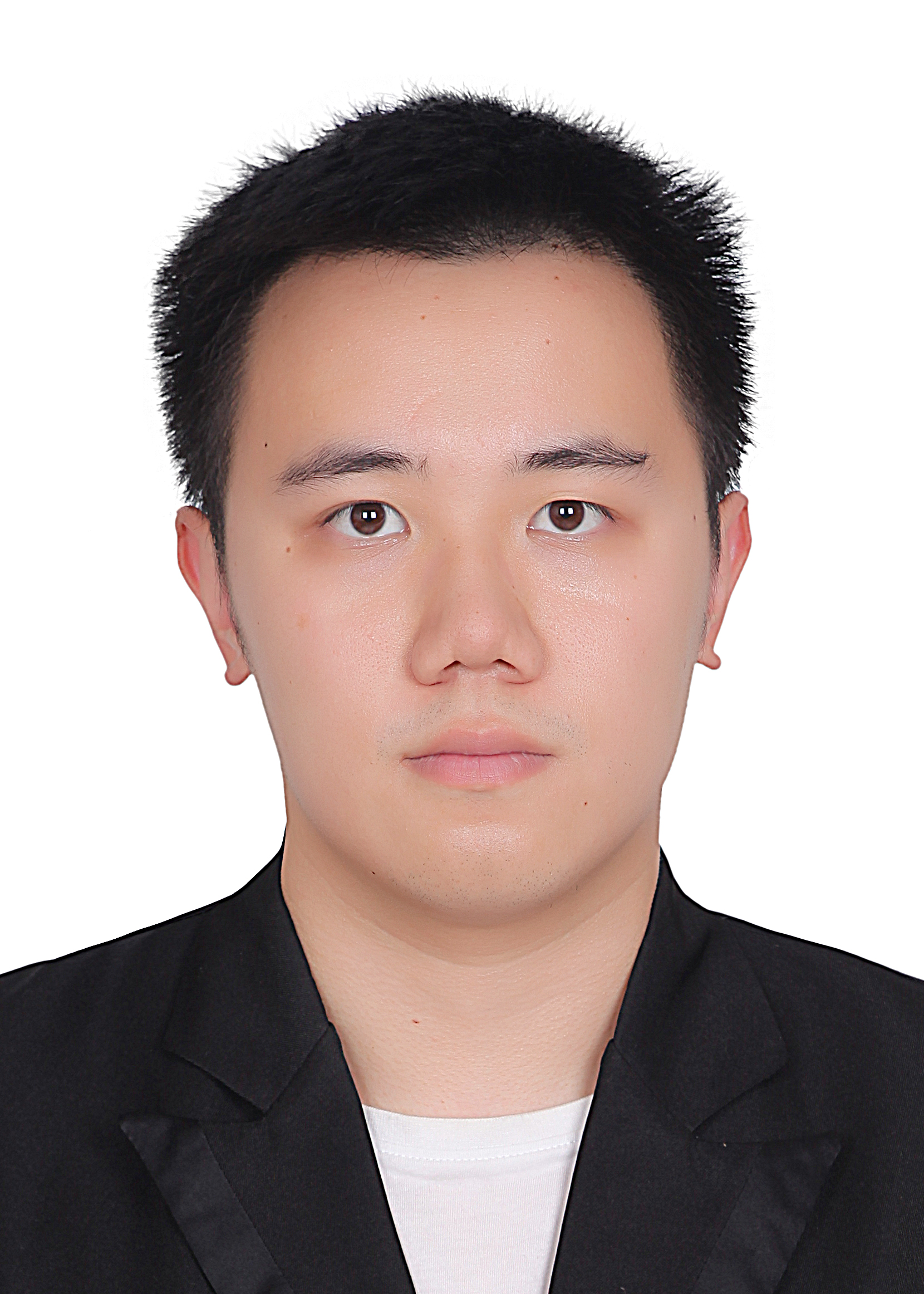}}]{Yipeng Liu}
received the B.S. degree in communication engineering from Shandong University, Qingdao, China, in 2019. He is currently working toward the Ph.D degree in information and communication engineering at Shanghai Jiao Tong University, Shanghai, China. His research interests include visual quality assessment and quality-related optimization.
\end{IEEEbiography}

\begin{IEEEbiography}
[{\includegraphics[width=1in,height=1.25in,clip,keepaspectratio]{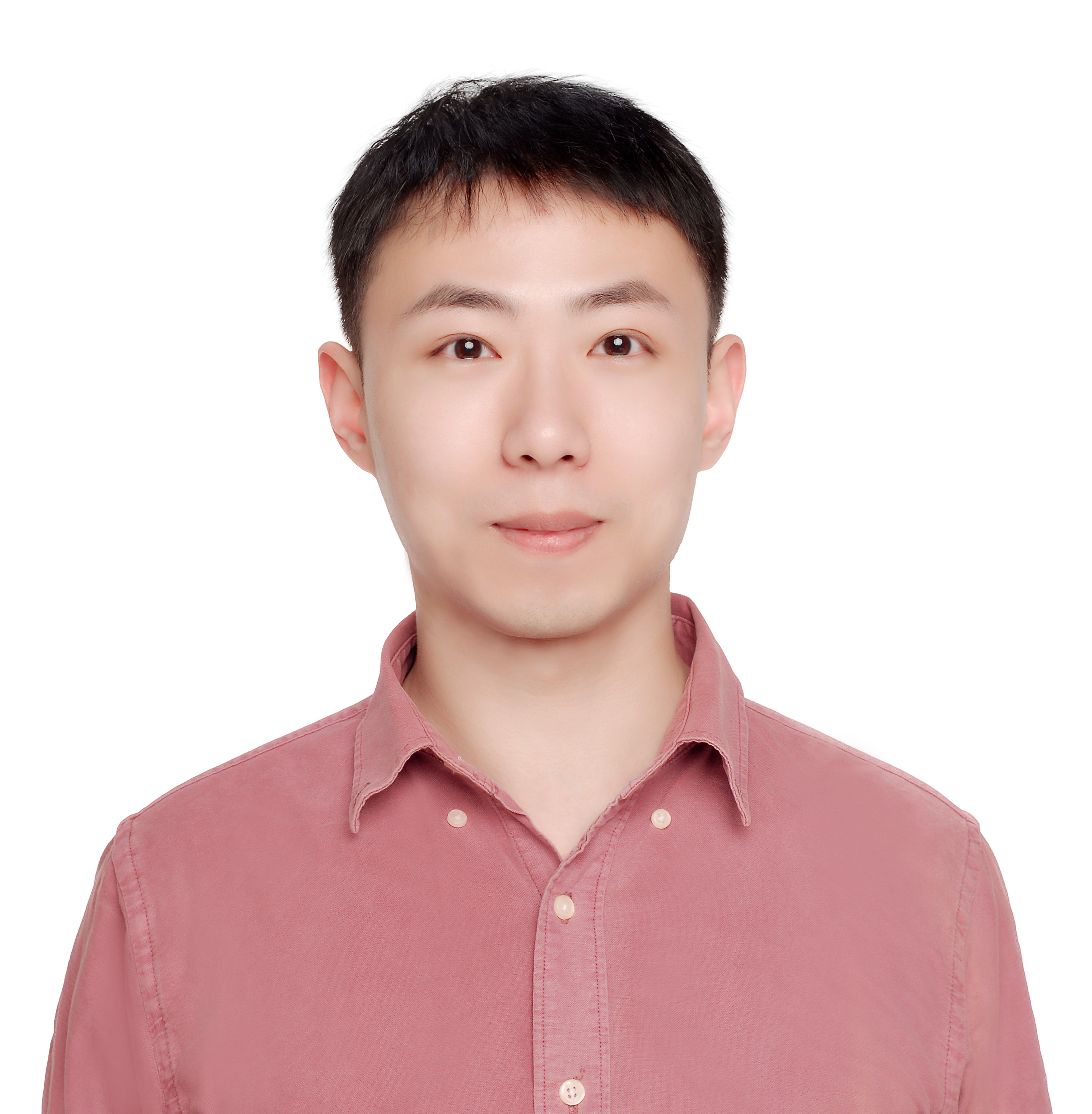}}]{Qi Yang} received the B.S. degree in communication engineering from Xidian University, Xi'an, China, in 2017, and Ph.D degree in information and communication engineering at Shanghai Jiao Tong University, Shanghai, China, 2022. He worked as a researcher in Tencent MediaLab from 2022 to 2024. Now, he joins University of Missouri–Kansas City as a post doctoral researcher. He has published more than 20 conference and journal articles, including TPAMI, TVCG, TMM, ACM TOMM, CVPR, ACM MM, ICME, etc. He is also an active member in standard organizations, including MPEG, AOMedia, and AVS. His research interests include image processing, 3D point cloud quality assessment and reconstruction, 3D mesh quality assessment and compression, and volumetric video quality assessment.
\end{IEEEbiography}

\begin{IEEEbiography}
[{\includegraphics[width=1in,height=1.25in,clip,keepaspectratio]{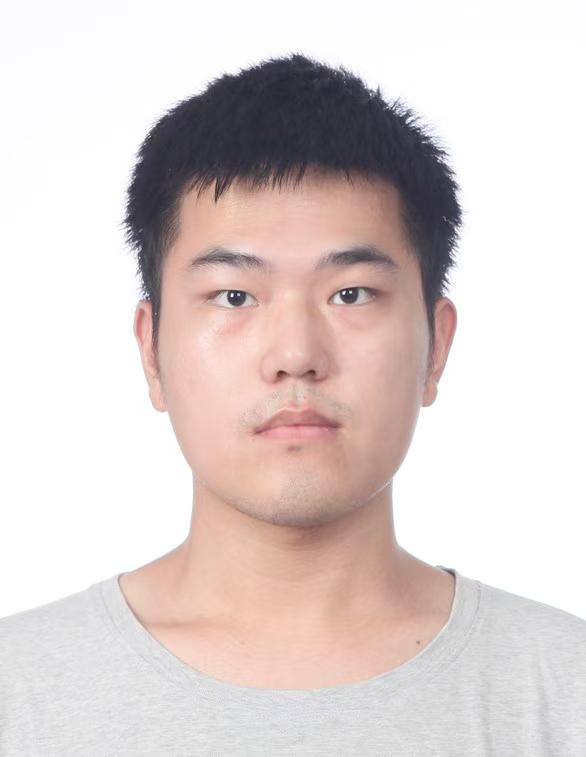}}]{Yujie Zhang}
received the B.S. degree in electronic and information engineering from Beihang University, Beijing, China, in 2020. He is currently working toward the PhD degree in information and communication engineering with Shanghai Jiao Tong University, Shanghai, China. His research interests include 3D quality assessment, geometry processing and novel view synthesis.
\end{IEEEbiography}

\begin{IEEEbiography}
[{\includegraphics[width=1in,height=1.25in,clip,keepaspectratio]{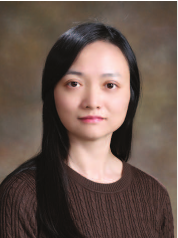}}]{Yiling Xu}
received the B.S., M.S., and Ph.D degrees from the University of Electronic Science and Technology of China, in 1999, 2001, and 2004 respectively. From 2004 to 2013, she was a senior engineer with the Multimedia Communication Research Institute, Samsung Electronics Inc., South Korea. She joined Shanghai Jiao Tong University, where she is currently a professor in the areas of multimedia communication, 3D point cloud compression and assessment, system design, and network optimization. She is the associate editor of the IEEE Transactions on Broadcasting. She is also an active member in standard organizations, including MPEG, 3GPP, and AVS.
\end{IEEEbiography}

\begin{IEEEbiography}
[{\includegraphics[width=1in,height=1.25in,clip,keepaspectratio]{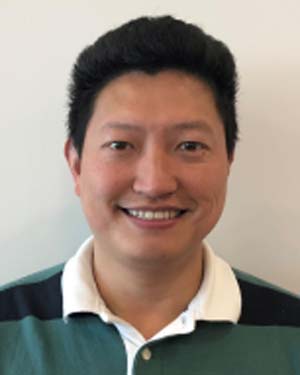}}]{Le Yang} received the B.S. degree in telecommunication engineering and M.S. degree in signal and information processing from the University of Electronic Science and Technology of China (UESTC), in 2000 and 2003, respectively, and the Ph.D degree in electrical and computer engineering (ECE) from the University of Missouri, Columbia, MO, in 2010. From 2003 to 2004, he was lecturer with UESTC. From 2004 to 2006, he was a research assistant with the University of Victoria (UVic), working on diversity techniques, performance analysis and ultra wideband (UWB) communications. From 2006 to 2007, he was with the McMaster University, working on the computational modelling of auditory cortex for studying the neural substrates of tinnitus. From 2011 to 2018, he was an Associate Professor with the Jiangnan University, Wuxi, China. In 2019, he joined the University of Canterbury, Christchurch New Zealand, as a senior lecturer with the Department of ECE. His research interests include statistical signal processing with applications to localization and tracking, machine learning, and engineering
optimization.
\end{IEEEbiography}

\begin{IEEEbiography}
[{\includegraphics[width=1in,height=1.25in,clip,keepaspectratio]{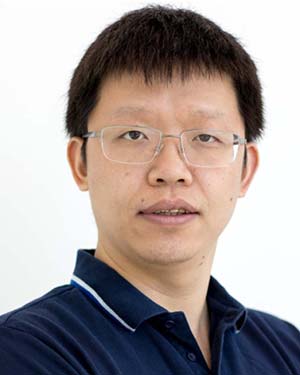}}]{Xiaozhong Xu} received the B.S. degree in electronics engineering from
Tsinghua University, Beijing, China, the M.S. degree in electrical and computer engineering from the
Polytechnic School of Engineering, New York University, NY, USA, and the Ph.D. degree in electronics engineering from Tsinghua University. He has been a Senior Principal Researcher and a Senior Manager of multimedia standards with Tencent
Media Lab, since 2017. He was with MediaTek as a Senior Staff Engineer and the Department Manager of Multimedia Technology Development, from 2013 to 2017. Prior to that, he worked for Zenverge (acquired by NXP in 2014), a semiconductor company focusing on multi-channel video transcoding ASIC design,
from 2011 to 2013. He also held technical positions at Thomson Corporate Research (now Technicolor) and Mitsubishi Electric Research Laboratories. His research interests include the general area of multimedia, including video, image and volumetric data coding, processing, and transmission. He has
been an active participant in various multimedia standardization activities for over 15 years. He was a recipient of the ISO\&IEC Excellence Award, the Technology Lumiere Award, and the Science and Technology Award from
China Association for Standardization. 
\end{IEEEbiography}

\begin{IEEEbiography}
[{\includegraphics[width=1in,height=1.25in,clip,keepaspectratio]{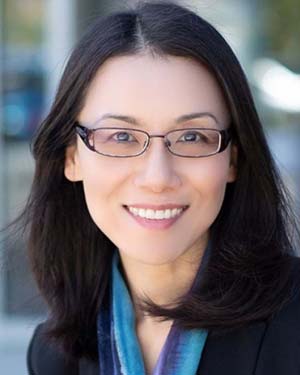}}]{Shan Liu} received the B.Eng.
degree in electronic engineering from Tsinghua University and the M.S. and Ph.D. degrees in electrical engineering from the University of Southern California. She was the formerly Director of the Media Technology Division, MediaTek, USA.
She was also formerly with MERL and Sony. She is a Distinguished Scientist and the General Manager with Tencent. She holds more than 600 granted U.S. patents and has published more than 100 peer-reviewed articles and one book. Her
research interests include audio-visual, volumetric, immersive, and emerging multimedia compression, intelligence, transports, and systems. She has been a long-time contributor to international standardization with many technical
proposals adopted into various standards, such as VVC, HEVC, OMAF, DASH, MMT, and PCC; and served as the Project Editor of ISO/IEC|ITUT H.266/VVC Standard. She was a recipient of the ISO\&IEC Excellence Award, the Technology Lumiere Award, the USC SIPI Distinguished Alumni Award, and two-time IEEE TRANSACTIONS ON CIRCUITS AND SYSTEMS FOR VIDEO TECHNOLOGY Best AE Award. She serves as an Associate Editor-in-Chief for IEEE TRANSACTIONS ON CIRCUITS AND SYSTEMS FOR VIDEO TECHNOLOGY and the Vice Chair for the IEEE Data Compression Standards Committee. She also serves and has served on a few other boards
and committees.
\end{IEEEbiography}

\end{document}